\documentclass[prb,twocolumn,showpacs,preprintnumbers,amsmath,aps]{revtex4}
\usepackage{graphicx,hyperref}% Include figure files
\usepackage{bm}% bold math
\usepackage{subfigure}% bold math

\newcommand\vect[1]{ \boldsymbol{ #1}}
\def\nbC{{\mathchoice {\setbox0=\hbox{$\displaystyle\rm C$}%
\hbox{\hbox to0pt{\kern0.4\wd0\vrule height0.9\ht0\hss}\box0}}
{\setbox0=\hbox{$\textstyle\rm C$}\hbox{\hbox
to0pt{\kern0.4\wd0\vrule height0.9\ht0\hss}\box0}}
{\setbox0=\hbox{$\scriptstyle\rm C$}\hbox{\hbox
to0pt{\kern0.4\wd0\vrule height0.9\ht0\hss}\box0}}
{\setbox0=\hbox{$\scriptscriptstyle\rm C$}\hbox{\hbox
to0pt{\kern0.4\wd0\vrule height0.9\ht0\hss}\box0}}}}
\def\nbQ{{\mathchoice {\setbox0=\hbox{$\displaystyle\rm
Q$}\hbox{\raise
0.15\ht0\hbox to0pt{\kern0.4\wd0\vrule height0.8\ht0\hss}\box0}}
{\setbox0=\hbox{$\textstyle\rm Q$}\hbox{\raise
0.15\ht0\hbox to0pt{\kern0.4\wd0\vrule height0.8\ht0\hss}\box0}}
{\setbox0=\hbox{$\scriptstyle\rm Q$}\hbox{\raise
0.15\ht0\hbox to0pt{\kern0.4\wd0\vrule height0.7\ht0\hss}\box0}}
{\setbox0=\hbox{$\scriptscriptstyle\rm Q$}\hbox{\raise
0.15\ht0\hbox to0pt{\kern0.4\wd0\vrule height0.7\ht0\hss}\box0}}}}
\def\nbT{{\mathchoice {\setbox0=\hbox{$\displaystyle\rm
T$}\hbox{\hbox to0pt{\kern0.3\wd0\vrule height0.9\ht0\hss}\box0}}
{\setbox0=\hbox{$\textstyle\rm T$}\hbox{\hbox
to0pt{\kern0.3\wd0\vrule height0.9\ht0\hss}\box0}}
{\setbox0=\hbox{$\scriptstyle\rm T$}\hbox{\hbox
to0pt{\kern0.3\wd0\vrule height0.9\ht0\hss}\box0}}
{\setbox0=\hbox{$\scriptscriptstyle\rm T$}\hbox{\hbox
to0pt{\kern0.3\wd0\vrule height0.9\ht0\hss}\box0}}}}
\def\nbS{{\mathchoice
{\setbox0=\hbox{$\displaystyle     \rm S$}\hbox{\raise0.5\ht0%
\hbox to0pt{\kern0.35\wd0\vrule height0.45\ht0\hss}\hbox
to0pt{\kern0.55\wd0\vrule height0.5\ht0\hss}\box0}}
{\setbox0=\hbox{$\textstyle        \rm S$}\hbox{\raise0.5\ht0%
\hbox to0pt{\kern0.35\wd0\vrule height0.45\ht0\hss}\hbox
to0pt{\kern0.55\wd0\vrule height0.5\ht0\hss}\box0}}
{\setbox0=\hbox{$\scriptstyle      \rm S$}\hbox{\raise0.5\ht0%
\hboxto0pt{\kern0.35\wd0\vrule height0.45\ht0\hss}\raise0.05\ht0%
\hbox to0pt{\kern0.5\wd0\vrule height0.45\ht0\hss}\box0}}
{\setbox0=\hbox{$\scriptscriptstyle\rm S$}\hbox{\raise0.5\ht0%
\hboxto0pt{\kern0.4\wd0\vrule height0.45\ht0\hss}\raise0.05\ht0%
\hbox to0pt{\kern0.55\wd0\vrule height0.45\ht0\hss}\box0}}}}
\def\nbZ{{\mathchoice {\hbox{$\sf\textstyle Z\kern-0.4em Z$}}
{\hbox{$\sf\textstyle Z\kern-0.4em Z$}}
{\hbox{$\sf\scriptstyle Z\kern-0.3em Z$}}
{\hbox{$\sf\scriptscriptstyle Z\kern-0.2em Z$}}}}
\begin{document}

\title{Non-Perturbative Functional Renormalization Group for Random Field Models and Related Disordered Systems. I: Effective Average Action Formalism}

\author{Gilles Tarjus} \email{tarjus@lptl.jussieu.fr}
\affiliation{LPTMC, CNRS-UMR 7600, Universit\'e Pierre et Marie Curie,
bo\^ite 121, 4 Pl. Jussieu, 75252 Paris c\'edex 05, France}

\author{Matthieu Tissier} \email{tissier@lptl.jussieu.fr}
\altaffiliation{Present address: Instituto de F\'\i{}sica, Facultad de
ingener\'\i{}a, Universidad de la Rep\'ublica, J.H. y Reissig 565,
11000 Montevideo, Uruguay}
\affiliation{LPTMC, CNRS-UMR 7600, Universit\'e Pierre et Marie Curie,
bo\^ite 121, 4 Pl. Jussieu, 75252 Paris c\'edex 05, France}

\date{\today}

\begin{abstract}
  We have developed a nonperturbative functional renormalization group
  approach for random field models and related disordered systems for
  which, due to the existence of many metastable states, conventional
  perturbation theory often fails. The approach combines an exact
  renormalization group equation for the effective average action with
  a nonperturbative approximation scheme based on a description of the
  probability distribution of the renormalized disorder through its
  cumulants. For the random field $O(N)$ model, the minimal truncation
  within this scheme is shown to reproduce the known perturbative
  results in the appropriate limits, near the upper and lower critical
  dimensions and at large number $N$ of components, while providing a
  unified nonperturbative description of the full $(N,d)$ plane, where
  $d$ is the spatial dimension.
\end{abstract}


\maketitle

\section{Introduction}\label{sec:introduction}

The effect of quenched disorder on the long-distance physics of
many-body systems largely remains an unsettled question despite
decades of intensive research. Ongoing controversies persist for
instance on the equilibrium and out-of-equilibrium behavior of spin
glasses and systems coupled to a random
field.\cite{young98,dedominicis06} Even though progress has been made,
it has so far proven difficult to construct a proper renormalization
group (RG) approach providing a description of ordering transitions
and criticality in these systems. A technical reason for this
unsatisfactory situation is that quenched disorder makes the system
intrinsincally inhomogeneous and that one should in principle follow
the renormalization of the whole probability distribution of the
disorder. A physical reason is that the presence of disorder and of
the resulting spatial inhomogeneity lead, for at least some range of
the control parameters, to multiple ``metastable states''.  (At this
point we use the term ``metastable state'' in a loose acceptance to
describe configurations that minimize some energy or free-energy,
action or effective action in field-theoretical terminology, but
differ from the true ground state.)  How such metastable states evolve
upon coarse-graining under RG then represents the central issue: at
large lengthscale, their influence could vanish, leaving only benign
signatures in the thermodynamics, or else it could modify the critical
behavior of the system, the nature of its phases, and, often in an
even more spectacular way, the relaxation and out-of-equilibrium
dynamical properties.

A well known example of the kind of puzzles associated with quenched
disorder and metastable states is the failure of the so-called
``dimensional reduction'' property in the random field Ising model
(RFIM).\cite{aharony76,grinstein76,young77,nattermann98} Standard
perturbation theory predicts to all orders that the critical behavior
of the RFIM in dimension $d$ is the same as that of the pure Ising
model, \textit{i.e.}, in the absence of random field, in two
dimensions less, $d-2$. The property has been shown in a compact and
elegant manner by Parisi and Sourlas\cite{parisi79} by means of a
supersymmetric formalism. However, dimensional reduction predicts a
lower critical dimension for ferromagnetism in the RFIM of $d_{lc}=3$,
in contradiction with rigorous results.\cite{imbrie84,bricmont87} The
dimensional reduction property must therefore break down in low enough
dimension. The supersymmetric approach gives a hint at the origin of
the breakdown, which appears to be related, yet in a somewhat obscure
way, to the presence of multiple metastable states\cite{parisi84b} (in
this case, local minima of the Hamiltonian).

Over the years, and on top of numerous computer simulations and scarce
exact analytical results, theoretical approaches have been devised to
cope with disordered systems characterized by multiple metastable
states, such as spin glasses and random field models.\cite{young98} To
list the main ones, we mention:

(i) phenomenological approaches such as the heuristic domain-wall
arguments\cite{imry75,bray85} and the ``droplet''
description,\cite{bray84, fisher88, fisher88b} in which one directly
focuses on rare excitations and the associated low-energy metastable
states;

(ii) mean-field theories, combined with the replica formalism in order
to handle the average over disorder; for models with spin-glass
ordering, the potentially dramatic effect of the metastable states is
captured through a spontaneous breaking of the replica
symmetry;\cite{mezard87,dotsenko01,young98,dedominicis06}

(iii) specific RG techniques for low-dimensional ($d=1,2$) systems, as
for instance the Coulomb gas RG approach for two-dimensional
disordered $XY$ models\cite{cardy82,carpentier98} or real space RG for
strongly disordered one-dimensional
systems;\cite{dasgupta80,fisher94,igloi05}

(iv) the perturbative functional RG for energy-dominated disordered
models considered in the vicinity of a critical dimension at which the
fundamental fields are
dimensionless;\cite{fisher85,fisher86,balents93,narayan93,ledoussal04,nattermann92}
one must then follow the flow of a whole function, an appropriate
renormalized cumulant of the disorder. As shown first by
Fisher\cite{fisher86} for an elastic manifold pinned by a random
potential, the long-distance physics is controlled by a
zero-temperature fixed point at which the renormalized cumulant is a
nonanalytic function of the fields, with the nonanalyticity encoding
the effect of the many metastable states at zero temperature.

All these approaches, however, are either questionable or not easily
generalizable: on the one hand, the phenomenological approaches lack
rigorous foundations and the relevance of mean-field descriptions to
finite-dimensional systems is, to say the least, far from garanteed;
on the other hand, the perturbative functional RG becomes extremely
complex, and soon untractable in practice for random field systems when going beyond one-loop
calculations;\cite{ledoussal06,tissier06,tissier06b} moreover, it does
not allow one to study the RFIM (as for specific RG techniques, they
are not extendable by construction).

The purpose of the present work, described here and in a companion
paper\cite{tarjus07_2} is therefore to propose a general theoretical
framework that leads to a consistent description of the equilibrium
behavior of the random field models and related disordered
systems. To achieve this, we rely on a version of Wilson's continuous
RG via momentum shell integration.\cite{wilson74} Under various
terminologies, ``Exact RG'', ``Functional RG'', and ``Nonperturbative
RG'', it has been developed in the past 15 years to become a powerful
method for investigating both universal and nonuniversal properties in
Statistical Physics and Quantum Field
Theory.\cite{morris98b,bagnuls01,berges02,delamotte03,pawlowski05} The
approach is ``exact'' in the sense that the RG flow associated with
the progressive account of the field fluctuations over larger and
larger lengthscales is described through an exact functional
differential equation. It is ``functional'' because through the exact
equation, one follows the flow of an infinite hierarchy of functions
of the fields in place of simply coupling constants. It is
``nonperturbative'' (beyond the mere tautology that an exact
description automatically includes all perturbative as well as
nonperturbative effects) because it lends itself to efficient
approximation schemes that are able to capture genuine nonperturbative
phenomena:\cite{berges02} to name a few, in the case of the standard
$O(N)$ scalar model, (numerically) tractable approximations describe
the Kosterlitz-Thouless transition of the $XY$ model in $d=2$, known
to be associated with the binding/unbinding of topological defects
(vortices), as well as the convexity property of the thermodynamic
potential in case of spontaneous symmetry breaking, recovered in other
treatments through nonperturbative configurations like instantons.

To study the problem at hand, we combine the ideas of the perturbative
functional RG for disordered systems with the general formalism of the
exact/functional/nonperturbative RG. In the following, we shall denote
our approach \textit{nonperturbative functional RG} (NP-FRG). It
provides a framework to study both perturbative and nonperturbative effects in any
spatial dimension $d$ and for any number of components of the
fundamental fields, $N$. We exclude from the scope of the present
series of articles relaxation and out-of-equilibrium dynamic
phenomena, as well as spin glass ordering. We also postpone to a
forthcoming publication the development of the NP-FRG in a superfield
formalism able to directly address the failure of supersymmetry in
connection with that of dimensional reduction. Short versions of the
present work have appeared in Refs.~[\onlinecite{tarjus04,tissier06}].

The present paper is organized as follows.

In section II we present the models and the formalism. We first
introduce the models and discuss their physical relevance and the main
open questions. From the corresponding replica field theories, we then
derive the exact RG equation for the effective average action, which
is the generating functional of the one-particle irreducible
correlation functions at the running scale. We next relate the replica
formalism, in which the replica symmetry is explicitly broken through
the application of sources, to the cumulants of the renormalized
disorder. We close the section by writing down the exact RG flow
equations for these cumulants.

In section III, we introduce a systematic nonperturbative
approximation scheme. After first discussing the symmetries of the
problem and the way to implement them in the effective average action
formalism, we introduce the nonperturbative truncation scheme of the
exact RG equation: it relies on (i) an expansion in cumulants of the
disorder and (ii) a well tested approximation of the nonperturbative
RG, the ``derivative expansion'', which uses the fact that the
relevant physics is dominated by long wavelength modes to perform an
expansion in the number of spatial derivatives of the fundamental
fields. Finally, we detail the minimal truncation that we use in our
numerical investigation of the random field $O(N)$ model (RF$O(N)$M).

In section IV, we specialize the formalism to the study of the
RF$O(N)$M. We introduce the scaling dimensions suitable to a search
for the putative zero-temperature fixed point controlling the ordering
transition. We first consider the case of the RFIM and then extend our
description to the RF$O(N)$M. With the help of these dimensions, the
RG flow equations are then cast in a scaled form. We also briefly
comment on possible application to other disordered systems.

We next discuss in section V an important property of the truncations
previously described: because of the one-loop structure of the exact
flow equations and of the appropriate choice of the approximations,
one recovers the perturbative results both near the upper critical
dimension, $d_{uc}=6$, and in the $N\rightarrow \infty$ limit of the
RF$O(N)$M. Even more interestingly, we also show that our minimal
truncation near the lower critical dimension for ferromagnetism of the
RF$O(N>1)$M, $d_{lc}=4$, reduces to the perturbative functional RG
result (at one loop) obtained from the nonlinear sigma version of the
model.\cite{fisher85} To the least, the truncated NP-FRG thus provides a
nonperturbative interpolation in the whole $(N,d)$ plane of the known
perturbative results near $d=4$, $d=6$, as well as $N\rightarrow
\infty$.

Finally, the presentation and discussion of the results obtained for
the RF$O(N)$M within the present NP-FRG approach will be described in
the companion paper.\cite{tarjus07_2}

\section{Models and formalism}
\label{sec_models}

\subsection{Models}

We focus on the equilibrium, long-distance behavior of a class of
disordered models in which $N$-component classical variables with
$O(N)$ symmetric interactions are coupled to a random field. Depending
on whether the coupling is linear or bilinear, the models belong to
the ``random field'' (RF) or the ``random anisotropy'' (RA)
subclasses. Such models with $N=1, 2$, or $3$ are relevant to describe
a variety of systems encountered in condensed matter physics or
physical chemistry. To name a few, one can mention dilute
antiferromagnets in a uniform magnetic field, \cite{belanger98}
critical fluids and binary mixtures in aerogels (both systems being
modelled by the $N=1$ RF Ising model),\cite{brochard83,degennes84,pitard95}
vortex phases in disordered type-II superconductors (described in
terms of an elastic glass model whose simplest version is the $N=2$ RF
$XY$ model),\cite{giamarchi98,blatter94,nattermann00} amorphous magnets, such as
alloys of rare-earth compounds,\cite{harris73,dudka05} and nematic liquid
crystals in disordered porous media (described by $N=2$ or $N=3$ RA
models).\cite{feldman01}

Other related models can be described as well within the same
formalism, but will only be alluded to: the ``random elastic'' model
describing an elastic system, such as an interface or a vortex lattice,
pinned by the presence of impurities; the ``random temperature'' model
associated with impurity-generated bond or site dilution in a
ferromagnetic Ising model. For reasons that will become clear further
down in this section, we exclude from the present study spin glass
ordering and we rather concentrate on ferromagnetic ordering (in which the $O(N)$ symmetry is spontaneously broken) or
``quasi-ordering'' (phases with quasi-long range order).

Our starting point is the field theoretical (coarse-grained)
description of the systems in terms of an $N$-component scalar field
$\vect \chi(\vect x)$ in a $d$-dimensional space and an effective
Hamiltonian, or bare action,

\begin{equation}
\begin{aligned}
\label{eq_ham_dis}
S[\vect \chi;\vect h, \vect \tau]= &\int_{\vect x} \bigg\{ \frac
{1}{2} \sum_{\mu=1}^{N} \left( \vert \vect \partial \chi^{\mu}(\vect
  x)\vert^2 + \tau \chi^{\mu}(\vect x)^2 \right) + \\& \frac{u}{4!} 
\left(\sum_{\mu=1}^{N}  \chi^{\mu}(\vect x)^2\right) ^2 -
\sum_{\mu=1}^{N} h^{\mu}(\vect x) \chi^{\mu}(\vect x) \\& -
\sum_{\mu,\nu=1}^{N} \tau^{\mu\nu}(\vect x) \chi^{\mu}(\vect x)
\chi^{\nu}(\vect x)\bigg\} ,
\end{aligned}
\end{equation}
where $ \int_{\vect x} \equiv \int d^d x$ and the superscript $\mu$
spans the $N$ components of the field; $\vect{h}(\vect{x})$ is a
random magnetic field and $\vect{\tau}(\vect{x})$ a second-rank random
anisotropy tensor, which are both taken for simplicity (see also
discussion below) with gaussian distributions characterized by zero
means and variances given by
\begin{equation}
  \label{eq_cumulant_h}
  \overline{h^\mu(\vect x)h^\nu(\vect y)}=\Delta\ \delta_{\mu\nu}\ \delta(\vect x-\vect y)
\end{equation}
\begin{equation}
  \label{eq_cumulant_tau}
\overline{\tau^{\mu\nu}(\vect x)\tau^{\rho\sigma}(\vect y)}=\frac{\Delta_{2}}{2} \left( \delta_{\mu\rho}\delta_{\nu\sigma}+\delta_{\mu\sigma}\delta_{\nu\rho}\right) \delta(\vect x-\vect y),
\end{equation}
where the overbar generically denotes the average over quenched
disorder. Higher-order random anisotropies could be included as
well. They will indeed be generated along the RG flow. However, for
symmetry reasons, when starting with only a second-rank, or more
generally an even-rank, random anisotropy, only even-rank anisotropies
are generated: this corresponds to what is called the random
anisotropy (RA) model. The model with a nonzero $\Delta$, for which
anisotropies of both odd and even ranks are generated under RG flow,
is the random field (RF) model.

The equilibrium properties of the model are obtained from the average
over disorder of the logarithm of the partition function,
\begin{equation}
  \label{eq_part_func}
\mathcal Z[\vect J]=\int\mathcal D \vect \chi  \exp \left(-S[\vect \chi;\vect h, \vect \tau] +   \int_{\vect x} \vect J(\vect x) \cdot \vect \chi(\vect x)\right) ,
\end{equation}
where $\vect J(\vect x)$ is a source linearly coupled to the
fundamental field and a (ultra-violet) momentum cutoff $\Lambda$,
associated with an inverse microscopic lengthscale such as a lattice
spacing, is implicitly considered in the functional integration over
the field. With this definition however, the partition function and
the corresponding thermodynamic potential $W[\vect J]=ln\mathcal
Z[\vect J]$ are still functionals of the random fields: $W[\vect
J]\equiv W[J;\vect h, \vect \tau]$. As is well known from the theory
of systems with quenched disorder, the thermodynamics is given by the
average over disorder of the ``free energy'', \textit{i.e.},

\begin{equation}
  \label{eq_averageW}
 \overline{W[\vect J]}=\overline{ln \mathcal Z[\vect J]}.
\end{equation}
Full information on the system, in particular an access to the
correlation (Green) functions of the field, requires knowledge of the
higher moments of $W[\vect J]$, viewed as a random
functional.\footnote{ An alternative to the description of the random
  functional by its moments or cumulants would be to directly consider
  its probability distribution. Unfortunately, this turns out to be
  extremely involved in the context of the present exact RG approach
  and for the models under consideration. For the $3-d$ hard-spin
  lattice model a numerical study of this kind has been carried out
  within a real-space RG description in the Migdal-Kadanoff
  approximation.\cite{falikov95} For simpler cases, this alternative
  approach has been followed: see for instance
  Ref.~[\onlinecite{carpentier98}].}  As will be discussed more
thoroughly further below, such information can be conveniently
extracted by using the replica formalism whose starting point is the
replacement of $ln \mathcal Z$ by the limit of $(\mathcal Z^n - 1)/n$
when $n$, the number of replicas of the original system, goes to
zero. Quite differently from the standard but controversial use of
this replica trick, in which the analytic continuation for $n<1$ opens
the possibility of a \textit{spontaneous} breaking of the replica
symmetry,\cite{mezard87} we will consider an \textit{a priori} more
benign procedure in which the symmetry between replicas is
\textit{explicitly} broken by the introduction of external sources
acting on each replica independently. This procedure will allow us to
generate the cumulant expansion of the disorder-dependent functional
$W[\vect J]$.

Within the replica formalism, the original problem is replaced by one
with $n$ replica fields $\{\vect \chi_a(\vect x)\}$, $a=1,2,\cdots,n$,
and the ``replicated action'', obtained after explicitly performing
the average over the disorder in the partition function, reads:
\begin{equation}
\label{eq_replicated_bare}
\begin{aligned}
S_{n}\left[\{\vect \chi_a\}\right ]&=  \int_{\vect x}\bigg\{\frac 1{2}\sum_{a=1}^n\big[\vert \partial \vect \chi_a(\vect x)\vert ^2+\tau \vert \vect \chi_a(\vect x) \vert^2+\\& \frac u{12}(\vert \vect \chi_a(\vect x)\vert^2) ^2\big]-\frac 1{2} \sum_{a,b=1}^n \big[\Delta \, \vect \chi_a(\vect x)\cdot \vect \chi_b(\vect x) \\& + \Delta_{2} (\vect \chi_a(\vect x)\cdot \vect \chi_b(\vect x))^{2} \big] \bigg\}
\end{aligned}
\end{equation}
with the corresponding partition function

\begin{equation}
\label{eq_replicated_partition}
\begin{aligned}
\mathcal Z_n \left[ \left\lbrace \vect J_a \right\rbrace  \right ] =  \int \prod_{a=1}^n  \mathcal D \vect \chi_a & \,  \exp \bigg( -S_{n}\left[ \left\lbrace \vect \chi_a \right\rbrace \right ] \\& +  \sum_{a=1}^n  \int_{\vect x} \vect J_a(\vect x)\cdot \vect \chi_a(\vect x) \bigg)
\end{aligned}
\end{equation}
where the linear sources $\vect J_a(\vect x)$, $a=1,2,\cdots,n$, act on each replica separately. Associated to this partition function is
the generating functional of the connected Green functions, $W_n
[\{\vect J_a\} ]=\ln \mathcal Z_n[\{\vect J_a\} ]$, and the effective
action, $\Gamma_n[\{\vect \phi_a\}]$, defined through a Legendre
transform:

\begin{equation}
\label{eq_replica_legendre_gamma}
\Gamma_n[\{\vect \phi_a \}] = - W_n[\{\vect J_a\}] + \sum_{a=1}^n \int_{\vect x} \vect J_a(\vect x)\cdot \vect \phi_a(\vect x),
\end{equation}
the fields $\{\vect \phi_a\}$ and the sources $\{\vect J_a\}$ being related by
\begin{subequations}
\begin{equation}
\label{eq_replica_legendre_phi}
\phi_a^\mu(\vect x)=\langle\chi_a^\mu(\vect x)\rangle=\frac{\delta W_n
  [\{\vect J_a\} ] }{\delta  J_a^\mu(\vect x)}
\end{equation}
where $\langle X\rangle$ represents the average of $X$ with the weight given in Eq.~(\ref{eq_replicated_partition}), and 
\begin{equation}
\label{eq_replica_legendre_J}
J_a^\mu(\vect x)=\frac{\delta \Gamma_n
  [\{\vect \phi_a\} ] }{\delta \phi_a^\mu(\vect x)}.
\end{equation}
\end{subequations}
The effective action is the generating functional of the one-particle
irreducible ($1-PI$) correlation functions or proper vertices.

The formalism we are about to describe also applies to extensions of
the replicated action of Eq.~(\ref{eq_replicated_bare}) that can be
cast in the form
\begin{equation}
\begin{split}
\label{eq_replicated_generalaction}
S_{n}\left[\{\vect \chi_a\}\right ]=  &\int_{\vect x}  \bigg\{\sum_{a=1}^n\big[\frac 1{2} \vert \partial \vect \chi_a(\vect x)\vert ^2+  U_{\Lambda}(\vect \chi_a(\vect x))\big]\\
&-\frac 1{2} \sum_{a,b=1}^n V_{\Lambda}(\vect \chi_a(\vect x), \vect \chi_b(\vect x)) + \cdots \bigg\},
\end{split}
\end{equation}
where the subscript $\Lambda$ recalls that the various terms are at
their bare value, defined at the microscopic scale $\Lambda$, and the
dots indicate possible functions involving higher numbers of
replicas. The functions $U_{\Lambda},V_{\Lambda},\cdots$ satisfy the
$O(N)$ symmetry as well as the $S_n$ permutational symmetry between
replicas. Eq.~(\ref{eq_ham_dis}) is obviously a special case of the
above expression, and higher-order anisotropies are included in a
$2$-replica term which is only function of $\vect \chi_a(\vect x)\cdot
\vect \chi_b(\vect x)$. RF and RA $O(N)$ models with nongaussian
distributions of the random fields and anisotropies are described by
terms involving higher number of replicas. (Note that the RA $O(N)$
model is defined as such for $N>1$; the Ising case, $N=1$, corresponds
to another model, the random temperature one introduced hereafter.)

Other disordered systems are also described by the form of the
replicated action in Eq.~(\ref{eq_replicated_generalaction}). For
instance, the random temperature model corresponds to
Eq.~(\ref{eq_replicated_generalaction}) with $U_\Lambda$ and
$V_\Lambda$ functions of the fields only through the $O(N)$ invariants
$\rho_a =\frac 1{2}\vert \vect \chi_a \vert^2 , \rho_b =\frac
1{2}\vert \vect \chi_b \vert^2$. In the RF, RA, and random temperature
models, the $1$-replica part of the bare action simply describes $n$
copies of the standard ferromagnetic $O(N)$ model without disorder.

The random elastic model is also a special case of
Eq.~(\ref{eq_replicated_generalaction}). However, contrary to the
models just discussed, the $1$-replica potential $U_\Lambda$ is absent
(or reduced to a purely quadratic term), so that there is no mechanism
triggering a paramagnetic-ferromagnetic phase transition. The
$2$-replica potential $V_\Lambda$, which is the second cumulant of a
random pinning potential, is now function of only the difference
between the two replica fields, $\vect \chi_a(\vect x) - \vect
\chi_b(\vect x)$. As a result, the model has an additional symmetry,
the statistical tilt symmetry,\cite{schultz88} which garantees that
the $1$-replica part of the action, including the kinetic term, is not
renormalized: the effective action has thus the same $1$-replica part
as the bare one. (Note that, as shown in Ref.~[\onlinecite{tissier06}]
and in the companion paper,\cite{tarjus07_2} the random elastic model,
albeit with an underlying periodicity, also emerges as a low-disorder
approximation of the RF and RA $XY$ ($N=2$) models.)

\subsection{Exact RG equation for the effective average action}

The exact RG in the effective average action
formalism\cite{wetterich93,morris98b,berges02} relates the bare action, here
Eq.~(\ref{eq_replicated_generalaction}), to the full effective action,
Eq.~(\ref{eq_replica_legendre_gamma}), through a progressive inclusion
of fluctuations of longer and longer wavelength. To do so, one
introduces an infrared regulator, characterized by a scale $k$, which,
in the functional integration leading to the partition function,
suppresses the contribution of the low-energy modes with momentum
$\vert \vect q \vert \lesssim k$ while including the high-energy modes
with $\vert \vect q \vert \gtrsim k$. After Legendre transformation,
this defines an ``effective average action'' at the running scale $k$,
$\Gamma_{k}$, which continuously interpolates between the microscopic
scale $k=\Lambda$, at which $\Gamma_{k=\Lambda}$ reduces to the bare
action, and the macroscopic one, $k=0$, at which $\Gamma_{k=0}$ equals
the full effective action.

More precisely in the present context, a ``mass-like'' quadratic term is added to the bare action, Eq.~(\ref{eq_replicated_generalaction}),
\begin{equation}
\label{eq_masslike}
\Delta \mathcal S_k[\{\vect \chi_a\}]=\frac 12 \sum_{a,b=1}^n \sum_{\mu,\nu=1}^N
\int_{\vect q} R_{k,ab}^{\mu\nu}(q^2)  \chi_a^\mu(-\vect q) \chi_b^\nu(\vect q),
\end{equation}
where $\int_{\vect q}\equiv \int d^dq/(2\pi)^d$;
$R_{k,ab}^{\mu\nu}(q^2)$ denotes infrared cutoff functions which, in
order to enforce that the additional term satisfies the same $O(N)$
and $S_n$ symmetries as the bare action (see above), must take the
following form:
\begin{equation}
\label{eq_replicatedregulator}
R_{k,ab}^{\mu\nu}(q^2) = \left(\widehat{R}_k(q^2)\delta_{ab}+\widetilde{R}_k(q^2) \right) \delta_{\mu\nu}.
\end{equation}

The cutoff functions $\widehat{R}_k(q^2)$ and $\widetilde{R}_k(q^2)$
are chosen such as to realize the decoupling of the low- and high-
momentum modes at the scale $k$: for this, they must decrease
sufficiently fast for large momentum $\vert \vect q \vert \gg k$ and
go to a constant value (a ``mass'') for small momentum $\vert \vect q
\vert \ll k$. The presence of an off-diagonal component
$\widetilde{R}_k(q^2)$ is somewhat unusual and will be discussed later
on.  The cutoff functions must also satisfy the two constraints that
(i) they go to zero when $k\to 0$, so that one indeed recovers the
full effective action with all modes accounted for, and (ii)
$\widehat{R}_k(q^2)$ diverges while $\widetilde{R}_k(q^2)$ stays
finite when $k\to \Lambda$, so that the effective average action does
reduce to the bare action. (In what follows we are only concerned with
the long-distance behavior of the models and do not pay attention to
microscopic details; we thus let $\Lambda$ go to $\infty$ in the
cutoff functions.) Different choices have been proposed and tested in
the recent literature. Standard choices for $\widehat{R}_k(q^2)$ are
of the form
\begin{equation}
\label{eq_scaledregulator}
\widehat{R}_k(q^2)=Z_k q^2 r(q^2/k^2)
\end{equation}
with $Z_k$ a field renormalization constant yet to be specified and
$r(y)=y^{-1}(1-y)\Theta(1-y)$,\cite{litim00} where $\Theta$ is the
Heaviside function, or $r(y)=(e^y -1)^{-1}$.\cite{wetterich93}

From the partition function $\mathcal Z_k\left[\{\vect J_a\}\right ]$
obtained from the bare action supplemented with the $k$-dependent
regulator, Eq.~(\ref{eq_masslike}), one defines the generating
functional of the Green functions $W_k [\{\vect J_a\} ]=\ln \mathcal
Z_k[\{\vect J_a\} ]$ and, through a Legendre transform, one has access
to the effective average action at the running scale $k$, $\Gamma_k$:
\begin{equation}
\label{eq_legendre_running_gamma}
\Gamma_k[\{\vect \phi_a \}] + W_k[\{\vect J_a\}] = \sum_{a=1}^n \int_{\vect x} \vect J_a(\vect x)\cdot \vect \phi_a(\vect x) - \Delta \mathcal S_k[\{\vect \phi_a\}]
\end{equation}
where the fields $\{\vect \phi_a\}$ and the sources $\{\vect J_a\}$ are related by the ($k$-dependent) expression
\begin{equation}
\label{eq_legendre_running_phi}
\phi_a^\mu(\vect x)=\langle\chi_a^\mu(\vect x)\rangle=\frac{\delta W_k
  [\{\vect J_a\} ] }{\delta  J_a^\mu(\vect x)}.
\end{equation}
The Legendre transform is slightly modified by the addition of the
last in Eq.~(\ref{eq_legendre_running_gamma}), which ensures that the
effective average action $\Gamma_k$ does reduce to the bare action at
the microscopic scale, with no contribution from the infrared
regulator. This addition does not change the behavior in the
$k\rightarrow 0$ limit since the regulator goes identically to
zero. Physically, and to use the language of magnetic systems, the
effective average action is a coarse-grained Gibbs free energy. It is
the generating functional of the $1-PI$ correlation functions from
which one can derive all Green functions of the modified system at the
scale $k$. Note that here and in the following we omit the subscript
$n$ associated to the number of replicas in order to simplify the
notations.

The evolution of the effective average action with the infrared cutoff
$k$ is governed by an exact flow equation,

\begin{equation}
\label{eq_erg}
\partial_k\Gamma_k\left[\{\vect \phi_a\}\right ]=
\dfrac{1}{2} \int_{\vect q} Tr \left\lbrace \partial_k \vect R_k(q^2) \left[\vect \Gamma _k^{(2)}+\vect R_k\right]_{\vect q,- \vect q}^{-1}\right\rbrace ,
\end{equation}
where the trace involves a sum over both replica indices and
$N$-vector components; $\vect R_k(q^2)$ is defined in
Eq.~(\ref{eq_replicatedregulator}) and $\vect \Gamma_k^{(2)}$ is the
tensor formed by the second functional derivatives of $\Gamma_k$ with
respect to the fields $\phi_a^\mu(\vect q)$:
\begin{equation}
\label{eq_secondderivative_gamma} 
 \left(\Gamma_k^{(2)}\right)_{ab}^{\mu\nu}(\vect q,\vect q')=
  \frac{\delta^2\Gamma_k}{\delta \phi_a^\mu(\vect q) \, \delta \phi_b^\nu(\vect q')}.
\end{equation}

The above RG flow equation is a complicated functional
integro-differential equation that cannot be solved exactly in
general; but, due to its one-loop structure and its reasonably
transparent physical content, it provides a convenient starting point
for nonperturbative approximation schemes.

At this point, it is quite clear to see why we have excluded spin
glass ordering from our considerations. The quadratic form of the
infrared regulator in Eq.~(\ref{eq_masslike}) suppresses the
fluctuations of the low-momentum modes of the fundamental fields
$\vect \chi_a$. Spin glass ordering on the other hand involves
fluctuations of composite fields, associated \textit{e.g.} to the
``overlap'' between different replicas.\cite{mezard87} Proper RG
treatment of such fluctuations implies to introduce a ``mass-like''
regulator for composite fields,\textit{ i.e.}, in the simplest case a
functional that is quartic in the fundamental fields instead of the
quadratic term used here. We do not consider this case in the present
work.

\subsection{Explicit replica symmetry breaking and cumulants of the renormalized disorder}

Among the technical difficulties encountered when making use of the
exact RG equation, Eq.~(\ref{eq_erg}), there is one which is specific
to disordered systems and to the present replica formalism: one must
invert the matrix $\vect \Gamma _{k,ab}^{(2)}+\vect R_{k,ab}$ for
arbitrary replica fields (since all replicas are different due to the
independently applied sources). Before delving into this problem, it
is worth giving some physical insight into the meaning of the explicit
replica symmetry breaking used here.

As discussed in section II-A, after full account of the fluctuations,
the bare disorder is renormalized to a full random (``free energy'')
functional $W[\vect J]$, which, to make its dependence on the bare
quenched disorder explicit, we now denote $W[\vect J;\vect h]$. This
random object can be characterized by the infinite set of its
cumulants, $W_1[\vect J_1], W_2[\vect J_1, \vect J_2],W_3[\vect J_1,
\vect J_2, \vect J_3],\cdots$, with
\begin{equation}
  \label{eq_cumW1}
W_1[\vect J_1]= \overline{W[\vect J_1;\vect h]}
\end{equation}
\begin{equation}
  \label{eq_cumW2}
W_2[\vect J_1, \vect J_2]= \overline{W[\vect J_1;\vect h]W[\vect J_2;\vect h]}-\overline{W[\vect J_1;\vect h]}\, \overline{W[\vect J_2;\vect h]},
\end{equation}
etc... The first cumulant $W_1$ gives access to the thermodynamics of
the system and the higher-order cumulants describe the distribution of
the renormalized disorder (we define, as in the bare action, a
disorder with zero mean). Note that by construction the cumulants are
invariant under permutations of their arguments.

The cumulants can be generated from an average involving copies, or
``replicas'', of the original disordered system, as follows:
\begin{equation}
\begin{split} 
 \label{eq_cumW}
 \overline{exp( \sum_{a=1} ^{n}W[\vect J_a; \vect h])} &= exp \left(
   W[\left\lbrace \vect J_a\right\rbrace ]\right) \\ = exp\bigg
 (&\sum_{a=1} ^{n} W_1[\vect J_a] +\dfrac{1}{2}\sum_{a,b=1}
 ^{n}W_2[\vect J_a, \vect J_b]\\+ \dfrac{1}{3!}&\sum_{a,b,c=1}
 ^{n}W_3[\vect J_a, \vect J_b, \vect J_c] + \cdots \bigg),
\end{split}
\end{equation}
where the $n$ copies have the \textit{same} bare disorder but are
coupled to different external sources. To fully characterize the
random functional $W[\vect J;\vect h]$, it is indeed important to
describe its cumulants for generic arguments, \textit{i.e.}, for
different sources. (Be aware that the subscripts $1,2, ...$ used to
denote the cumulants of $W$ should not be confused with the subscript
$n$ denoting the number of replicas in section II-A and omitted since:
here for instance, $W_1$ denotes the $1$-replica component,
corresponding to the first cumulant, whereas with the previous
notation $W_{n=1}$ is given by the sum of all cumulants with all there
arguments equal.)

A convenient trick to extract the cumulants with their full functional
dependence is to let the number of replicas be arbitrary and to view
the expansion in the right-hand side of Eq.~(\ref{eq_cumW}) as an
expansion in increasing number of ``free'', or unconstrained, sums
over replicas of the functional $W[\left\lbrace \vect J_a\right\rbrace
]$ defined below Eq.~(\ref{eq_replicated_partition}). The term of
order $p$ in the expansion is a sum over $p$ replica indices of a
functional depending exactly on $p$ replica sources, this functional
being precisely equal here to the $p$th cumulant of $W[\vect J;\vect
h]$. This procedure, which rests on an explicit breaking of the
replica symmetry and an analytic continuation to arbitrary numbers of
replicas (including the limit $n\rightarrow 0$ previously introduced),
is \textit{a priori} different from the standard use of replicas, in
which all sources are equal, and it avoids the delicate handling of a
spontaneous replica symmetry
breaking.\cite{mezard87,dotsenko01,young98,dedominicis06} It has been
used in a similar context by Le Doussal and
Wiese.\cite{ledoussal03,ledoussal04b} The practical implementation of
the expansion in free replica sums will be detailed in the next
subsection.

In our present NP-FRG approach however, the central object is the
effective action $\Gamma$, not $W$. The expansion of $\Gamma
[\left\lbrace \vect \phi_a\right\rbrace ]$ in increasing number of
free replica sums reads
\begin{equation}
\begin{split} 
 \label{eq_cumg}
\Gamma [\left\lbrace \vect \phi_a\right\rbrace ]= & \sum_{a=1} ^{n} \Gamma_1[\vect \phi_a] -\dfrac{1}{2}\sum_{a,b=1} ^{n}\Gamma_2[\vect \phi_a, \vect \phi_b]\\& + \dfrac{1}{3!}\sum_{a,b,c=1} ^{n}\Gamma_3[\vect \phi_a, \vect \phi_b, \vect \phi_c] + \cdots,
\end{split}
\end{equation}
where for later convenience we have introduced a minus sign for all
even terms of the expansion. $\Gamma [\left\lbrace \vect
  \phi_a\right\rbrace ]$ and $W[\left\lbrace \vect J_a\right\rbrace ]$
are related by a Legendre transform, so if one also expand the sources
$\vect J_a[\left\lbrace \vect \phi_f\right\rbrace ]$ (where we have
denoted $\left\lbrace \vect \phi_f\right\rbrace$ the $n$ replica
fields to avoid confusion in the indices) in increasing number of free
replica sums, one can relate the terms of the expansion of the
effective action to the cumulants of the random functional $W[\vect
J;\vect h]$. The relation is straighforward for the first terms, but
gets more involved as the order increases.

More precisely, $\Gamma_1[\vect \phi]$ is the Legendre transform of
$W_1[\vect J]$, namely,
\begin{equation}
\label{legendre_gamma_1}
\Gamma_1[\vect \phi ] = - W_1[\vect J] +  \int_{\vect x} \vect J(\vect x)\cdot \vect \phi(\vect x),
\end{equation}
with
\begin{equation}
\label{legendre_phi}
\phi^\mu(\vect x)=\frac{\delta W_1 [\vect J ] }{\delta  J^\mu(\vect x)},
\end{equation}
and the second-order terms is given by
\begin{equation}
  \label{eq_cumg2}
\Gamma_2[\vect \phi_1, \vect \phi_2] = W_2[\vect J[\vect \phi_1], \vect J[\vect \phi_2]],
\end{equation}
where $\vect J[\vect \phi]$ is the \textit{nonrandom} source defined
via the inverse of the Legendre transform relation in
Eq.~(\ref{legendre_gamma_1}), \textit{i.e.}, $ J^\mu[\vect \phi](\vect
x)=\delta \Gamma_1 [\vect \phi] / \delta \phi^\mu(\vect x)$. (Note
that $\vect J(x)$ introduced here differs from the source $\vect
J_a(x)$ introduced in equation (\ref{eq_replica_legendre_J}): through
the Legendre relations, the latter depends on all the fields
$\{\phi_a\}$ while the former depends on only one replica field.) The
above expression motivates our choice of signs for the terms of the
expansion in free replica sums of $\Gamma [\left\lbrace \vect
  \phi_a\right\rbrace ]$, Eq.~(\ref{eq_cumW}): $\Gamma_2[\vect \phi_1,
\vect \phi_2]$ is directly the second cumulant of $W[\vect J;\vect h]$
(with the proper choice of $\vect J[\vect \phi]$).

For the higher-order terms, one finds

\begin{equation}
\begin{split}
  \label{eq_cumg3}
\Gamma_3 [\vect \phi_1, &\vect \phi_2, \vect \phi_3]  =  - W_3[\vect J[\vect \phi_1], \vect J[\vect \phi_2],\vect J[\vect \phi_3]]  +  \\& \int_{\vect x \vect y} \bigg \{ W_{2, \vect x}^{(10)}[\vect J[\vect \phi_1], \vect J[\vect \phi_2]]\left( W_{1}^{(2)}[\vect J[\vect \phi_1]]\right)^{-1} _{\vect x\, \vect y}  \\& \times W_{2, \vect y}^{(10)}[\vect J[\vect \phi_1], \vect J[\vect \phi_3]] + perm (123)\bigg \} ,
\end{split}
\end{equation}
etc..., where $perm (123)$ denotes the two additional terms obtained by circular permutations of the fields $\vect \phi_1, \vect \phi_2, \vect \phi_3$ and where we have used the following short-hand notation:

\begin{equation}
 \label{eq_notation_partderiv_W1}
W_{1,\vect x_1 ... \vect x_p}^{(p)}[\vect J_1]=\frac{\delta^p W_1[\vect J_1]}{\delta J_1(\vect x_1)... \delta J_1(\vect x_p)},
\end{equation}
\begin{equation}
 \label{eq_notation_partderiv_W2}
\begin{split}
W_{2,\vect x_1 ... \vect x_p,\vect y_1 ... \vect y_q}^{(pq)} &[\vect J_1, \vect J_2]=\\ & \frac{\delta^{p+q} W_2[\vect J_1, \vect J_2]}{\delta J_1(\vect x_1)... \delta J_1(\vect x_p)\delta J_2(\vect y_1)... \delta J_2(\vect y_q)},
\end{split}
\end{equation}
etc. Note that for clarity the $O(N)$ indices have been omitted in the above expressions.

We point out that $\Gamma_p[\vect \phi_1, ..., \vect \phi_p]$ for $p\geq 3$ cannot be directly taken as the $p$th  cumulant of a physically accessible random functional, in particular not of the disorder-dependent Legendre transform of $W[\vect J;\vect h]$ (although it can certainly be expressed in terms of such cumulants of order equal or lower than $p$). In the following and by abuse of language, we will nonetheless generically call the $\Gamma_p$'s ``cumulants of the renormalized disorder'' (which is true for $p=2$).

In complement to the above picture and more specifically for random field systems, it is also interesting to introduce a renormalized random field (or random force) $ \breve{\vect h}[\vect \phi](\vect x)$ defined as the derivative of a random free-energy functional,
\begin{equation}
\label{def_ren_randomfield}
\vect \breve{h}[\vect \phi]^{\mu}(\vect x)=- \frac{\delta }{\delta \phi^\mu(\vect x)}\left(W[\vect J[\vect \phi];\vect h] - \overline{W[\vect J[\vect \phi];\vect h]} \right) ,
\end{equation}
and whose first moment is equal to zero by construction. It is easy to derive that its $p$th cumulant ($p\geq 2$) is given by the derivative with respect to $\vect \phi_1, ..., \vect \phi_p$ of $W_p[\vect J[\vect \phi_1], ...,\vect J[\vect \phi_p]]$, which can then be  related to derivatives of $ \Gamma_2, \Gamma_3,...$; for instance,
\begin{equation}
  \label{eq_cumhren2}
\overline{\breve{h}[\vect \phi_1](\vect x) \breve{h}[\vect \phi_2](\vect y)}= \Gamma_{2,\vect x \vect y}^{(11)}[\vect \phi_1,\vect \phi_2],
\end{equation}
where have used a short-hand notation similar to that of
Eqs.~(\ref{eq_notation_partderiv_W1},\ref{eq_notation_partderiv_W2})
and omitted the $N$-vector indices for simplicity. Terms of order $3$
and higher are again given by more complicated expressions.

We close this discussion by noticing that in the simpler case of the
random manifold model, $\Gamma_1$ and $W_1$ being trivial and
unrenormalized due to the statistical tilt symmetry (see above),
$\vect J [\vect \phi]$ has a simple explicit expression. For instance,
if the bare action has a quadratic $1$-replica term, $\Gamma_1[\vect
\phi]$ is equal to this quadratic functional and $\vect J [\vect
\phi]$ is a known linear functional of $\vect \phi$, which further
simplifies when considering uniform fields. This allows one to devise
ways to directly measure the second cumulant of the renormalized
disorder.\cite{ledoussal06b,middleton07} Nothing similar occurs in
random field and random anisotropy models: the thermodynamics of such
systems being highly nontrivial (with a phase transition and a
critical point), the expression of $\vect J [\vect \phi]$ is involved
and a priori unknown.

\subsection{Exact RG equations for the renormalized disorder cumulants}

The reasoning developed in the previous subsection can be applied to
the effective average action $\Gamma_k$ and its expansion in free
replica sums. As a results, Eqs.~(\ref{eq_cumW1}-\ref{eq_cumhren2})
can be extended to any running scale $k$. Yet, to make the expansion
in free replica sums an operational procedure, one needs be able to
perform systematic algebraic manipulations, as for instance the
inversion of the matrix appearing in the right-hand side of the exact
RG equation, Eq.~(\ref{eq_erg}). We detail here the method for
matrices depending on two replica indices, but functionals of the $n$
replica fields. Extension to higher-order tensors is presented in
Ref.~[\onlinecite{ledoussal04b}].

A generic matrix $A_{ab}[\left\lbrace \vect \phi_f\right\rbrace ]$,
where we have again denoted $\left\lbrace \vect \phi_f\right\rbrace$
the $n$ replica fields to avoid confusion in the indices, can be
decomposed as
\begin{equation}
\label{eq-genericdecompos}
A_{ab}[\left\lbrace \vect \phi_f\right\rbrace ]=\widehat{A}_{a}[\left\lbrace \vect \phi_f\right\rbrace ] \delta_{ab}+ \widetilde{A}_{ab}[\left\lbrace \vect \phi_f\right\rbrace ].
\end{equation}
In the above expression, it is understood that the second term
$\widetilde{A}_{ab}$ no longer contains any Kronecker symbol. Each
component can now be expanded in increasing number of free replica
sums,
\begin{equation}
\label{eq_widehatA}
\widehat{A}_{a}[\left\lbrace \vect \phi_f\right\rbrace ]=\widehat{A}^{[0]}[\vect \phi_a ]+\sum_{c=1} ^{n}\widehat{A}^{[1]}[\vect \phi_a \vert \vect \phi_c  ]+\cdots
\end{equation}
\begin{equation}
\label{eq_widetildeA}
\widetilde{A}_{ab}[\left\lbrace \vect \phi_f\right\rbrace ]=\widetilde{A}^{[0]}[\vect \phi_a, \vect \phi_b]+\sum_{c=1} ^{n}\widetilde{A}^{[1]}[\vect \phi_a, \vect \phi_b \vert  \vect \phi_c]+\cdots,
\end{equation}
where the superscripts in square brackets denote the order in the
expansion (and should not be confused with superscripts in parentheses
indicating partial derivatives).

As an illustration, the expansion of the matrix $\vect
\Gamma_{k}^{(2)}$ defined in Eq.~(\ref{eq_secondderivative_gamma})
reads, in terms of the expansion of effective average action itself,
\begin{equation}
\label{eq_widehatGamma}
\widehat{\vect \Gamma}_{k}^{(2)}[\left\lbrace \vect \phi_f\right\rbrace ]_{a}=\vect \Gamma_{k,1}^{(2)}[\vect \phi_a ]-\sum_{c=1} ^{n}\vect \Gamma_{k,2}^{(20)}[\vect \phi_a, \vect \phi_c  ]+\cdots
\end{equation}
\begin{equation}
\label{eq_widetildeGamma}
\begin{split}
  \widetilde{\vect \Gamma}_{k}^{(2)}[\left\lbrace \vect
    \phi_f\right\rbrace ]_{ab}=- &\vect \Gamma_{k,2}^{(11)} [\vect
  \phi_a, \vect \phi_b]+\\& \sum_{c=1} ^{n}\vect
  \Gamma_{k,3}^{(110)}[\vect \phi_a, \vect \phi_b, \vect
  \phi_c]+\cdots,
\end{split}
\end{equation}
where the permutational symmetry of the arguments of the
$\Gamma_{k,p}$'s has been used.

Algebraic manipulations on such matrices can be performed by
term-by-term identification of the orders of the expansions. For
instance, the inverse $\vect B=\vect A^{-1}$ of the matrix $\vect A$
can also be put in the form of Eq.~(\ref{eq-genericdecompos}) and its
components, $\widehat{B}_{a}$ and $\widetilde{B}_{ab}$, expanded in
number of free replica sums. The term-by-term identification of the
condition $\vect A \cdot \vect B= \vect 1$ leads to a unique
expression of the various orders, $\widehat{B}^{[p]}$ and
$\widetilde{B}^{[p]}$, of the expansion of $\vect B$ in terms of the
$\widehat{A}^{[q]}$'s and $\widetilde{A}^{[q]}$'s with $q\leq p$. The
algebra becomes rapidly tedious, but the first few terms are easily
derived:
\begin{equation}
\label{eq_widehatB_0}
\widehat{B}^{[0]}[ \vect \phi_1] = \widehat{A}^{[0]}[ \vect \phi_1]^{-1}
\end{equation}
\begin{equation}
\label{eq_widetildeB_0}
\widetilde{B}^{[0]}[ \vect \phi_1,\vect \phi_2] = - \widehat{B}^{[0]}[ \vect \phi_1]\widetilde{A}^{[0]}[ \vect \phi_1,\vect \phi_2] \widehat{B}^{[0]}[ \vect \phi_2]
\end{equation}
\begin{equation}
\label{eq_widehatB_1}
\widehat{B}^{[1]}[ \vect \phi_1 \vert \vect \phi_2] =  - \widehat{B}^{[0]}[ \vect \phi_1]\widehat{A}^{[1]}[ \vect \phi_1\vert \vect \phi_2] \widehat{B}^{[0]}[ \vect \phi_1]
\end{equation}
\begin{equation}
\label{eq_widetildeB_1}
\begin{split}
\widetilde{B} ^{[1]}[ &\vect \phi_1 , \vect \phi_2 \vert \vect \phi_3]  = - \widehat{B}^{[0]}[ \vect \phi_1] \bigg\{ \widetilde{A}^{[1]}[ \vect \phi_1,\vect \phi_2 \vert \vect \phi_3] \\& - \widetilde{A}^{[0]}[ \vect \phi_1,\vect \phi_3] \widehat{B}^{[0]}[ \vect \phi_3]\widetilde{A}^{[0]}[ \vect \phi_3,\vect \phi_2] \\&- \widehat{A}^{[1]}[ \vect \phi_1\vert \vect \phi_3] \widehat{B}^{[0]}[ \vect \phi_1] \widetilde{A}^{[0]}[ \vect \phi_1,\vect \phi_2] \\& - \widetilde{A}^{[0]}[ \vect \phi_1,\vect \phi_2] \widehat{B}^{[0]}[ \vect \phi_2] \widehat{A}^{[1]}[ \vect \phi_2\vert \vect \phi_3]\bigg\} \widehat{B}^{[0]}[ \vect \phi_2]
\end{split}
\end{equation}
etc.

We can apply the above procedure to the exact RG equation for the effective average action. For convenience, we introduce the modified propagator at the scale $k$,
\begin{equation}
\label{eq_propagator}
\vect P_k[\left\lbrace \vect \phi_f\right\rbrace ]=\left[\vect \Gamma _k^{(2)}+\vect R_k\right]^{-1},
\end{equation}
with
\begin{equation}
\label{eq_decomp_propagator}
\vect P_{k,ab}[\left\lbrace \vect \phi_f\right\rbrace ]= \widehat{\vect P}_{k,a}[\left\lbrace \vect \phi_f\right\rbrace ] \delta_{ab}+  \widetilde{\vect P}_{k,ab}[\left\lbrace \vect \phi_f\right\rbrace ],
\end{equation}
where $\widehat{\vect P}_{k,a}$ and $\widetilde{\vect P}_{k,ab}$ are
still tensors with respect to momenta and vector component
indices. Eq.~(\ref{eq_erg}) then leads to an infinite hierarchy of
flow equations for the cumulants of the renormalized disorder,
\begin{equation}
\label{eq_flow_Gamma1}
\begin{split}
\partial_k\Gamma_{k,1}\left[\vect \phi_1\right ]=
\dfrac{1}{2} \int_{\vect q} & \bigg \{ \partial_k (\widehat{R}_k(q^2)+ \widetilde{R}_k(q^2)) tr  \widehat{\vect P}_{k, \vect q\,- \vect q}^{[0]}\left[\vect \phi_1\right ]\\& +\partial_k \widehat{R}_k(q^2) tr \widetilde{\vect P}_{k, \vect q\,- \vect q}^{[0]}\left[\vect \phi_1,\vect \phi_1 \right ]\bigg \} ,
\end{split}
\end{equation}
\begin{equation}
\label{eq_flow_Gamma2}
\begin{split}
\partial_k& \Gamma_{k,2}\left[\vect \phi_1 ,\vect \phi_2\right ]= - 
\dfrac{1}{2} \int_{\vect q} \bigg \{ \partial_k (\widehat{R}_k(q^2)+ \widetilde{R}_k(q^2)) \\& tr \widehat{\vect P}_{k, \vect q\,- \vect q}^{[1]}\left[\vect \phi_1\vert \vect \phi_2\right ] + \partial_k \widehat{R}_k(q^2) tr \widetilde{\vect P}_{k, \vect q\,- \vect q}^{[1]}\left[\vect \phi_1,\vect \phi_1\vert \vect \phi_2 \right ] \\& + \partial_k \widetilde{R}_k(q^2) tr \widetilde{\vect P}_{k, \vect q\,- \vect q}^{[0]}\left[\vect \phi_1,\vect \phi_2 \right ]  + perm (12)\bigg \},
\end{split}
\end{equation}
and so on,  where $tr$ indicates a trace over $N$-vector components and $perm (12)$ denotes the expression obtained by permuting $\vect \phi_1$ and $\vect \phi_2$. (Some care is needed in the term by term identification in order to properly symmetrize the expressions and satisfy the permutational property of the various arguments of the cumulants.)

Expressing the higher-order terms $\widehat{\vect P}_{k}^{(p)}$ and
$\widetilde{\vect P}_{k}^{(p)}$ with $p\geq1$ only by means of
$\widehat{\vect P}_{k}^{[0]}$ and the derivatives of the $\Gamma
_{k,p}$'s and introducing the short-hand notation
$\widetilde{\partial}_k$ to indicate a derivative acting only on the
cutoff functions, \textit{i.e.}, $\widetilde{\partial}_k
\equiv \partial_k \widehat{R}_k\, \delta/\delta \widehat{R}_k
+ \partial_k \widetilde{R}_k \, \delta/\delta \widetilde{R}_k$,
Eq.~(\ref{eq_flow_Gamma2}) can be rewritten
\begin{equation}
\label{eq_flow_Gamma2_final}
\begin{split}
\partial_k& \Gamma_{k,2}\left[\vect \phi_1 ,\vect \phi_2\right ]= \dfrac{1}{2} \widetilde{\partial}_k \, Tr \bigg \{  \widehat{\vect P}_{k}^{[0]}\left[\vect \phi_1 \right ] ( \vect \Gamma _{k,2}^{(20)}\left[\vect \phi_1, \vect \phi_2 \right ] \\& - \vect \Gamma _{k,3}^{(110)}\left[\vect \phi_1, \vect \phi_1, \vect \phi_2 \right ]) + \widetilde{\vect P}_{k}^{[0]}\left[\vect \phi_1,\vect \phi_1 \right ] \vect \Gamma _{k,2}^{(20)}\left[\vect \phi_1,\vect \phi_2 \right ] \\& +\dfrac{1}{2} \widetilde{\vect P}_{k}^{[0]}\left[\vect \phi_1,\vect \phi_2 \right ] (\vect \Gamma _{k,2}^{(11)}\left[\vect \phi_1,\vect \phi_2 \right ] - \widetilde{R}_k \vect 1 )  + perm (12) \bigg \},
\end{split}
\end{equation}
and similarly for higher-order cumulants , where $ \vect 1_{\vect q \vect q'}^{\mu\nu}= (2\pi)^d \delta(\vect q + \vect q')\delta_{\mu\nu}$ and the trace $Tr$ is now over both momenta and and $N$-vector components; the modified propagators $\widehat{\vect P}_{k}^{[0]}$ and $\widetilde{\vect P}_{k}^{[0]}$ are explicitly given by

\begin{equation}
\label{eq_hatpropagator}
\widehat {\vect P}_{k}^{[0]}[ \vect \phi_1 ]=\left( \vect \Gamma _{k,1}^{(2)}[ \vect \phi_1 ]+\widehat R_k \vect 1\right) ^{-1},
\end{equation}
\begin{equation}
\label{eq_tildepropagator}
\widetilde {\vect P}_{k}^{[0]}[ \vect \phi_1,  \vect \phi_2 ]= \widehat {\vect P}_{k}^{[0]}[ \vect \phi_1 ](  \vect \Gamma _{k,2}^{(11)}[ \vect \phi_1,  \vect \phi_2 ]-\widetilde R_k \vect 1 ) \widehat {\vect P}_{k}^{[0]}[ \vect \phi_2 ]
\end{equation}

This provides a hierachy of exact RG equations for the cumulants of
the renormalized disorder (including the first one which leads to a
description of the thermodynamics). One should note that (i) the
cumulants are functional of the fields and contain full information on
the complete set of $1-PI$ correlation functions and (ii) the flow
equations are coupled, the $(p+1)$th cumulant appearing in the
right-hand side of the equation for the $p$th cumulant. As such these
RG equations remain untractable and their resolution requires
approximations.

\section{Nonperturbative approximation scheme}
\label{sec_NP}

\subsection{Symmetries in the effective average action formalism}

When writing the RG flow for the effective average action and when
devising an approximation scheme to solve it, one should as far as
possible make sure that the symmetries of the theory are not
explicitly violated at any scale. Such a requirement is easily
implemented as far as elementary symmetries, such as invariance by
translation and rotation in Euclidean space, $O(N)$ symmetry, and
$S_n$ replica permutational symmetry, are concerned: the infrared
regulator $\Delta \mathcal S_k$ added to the bare action must be
chosen such that it is invariant under the appropriate
transformations, which is indeed garanteed by the expressions in
Eqs.~(\ref{eq_masslike},\ref{eq_replicatedregulator}). The exact
effective average action at any scale $k$ then also possesses the
symmetries of the bare action, and one just had to be careful that the
truncations do not explicitly break the symmetries, which is easily
implemented.\cite{berges02}

A similar treatment can be applied to most additional symmetries of
the disordered systems under consideration. For instance, the
``statistical tilt symmetry'' of the random manifold model is easily
extended to a $k$-dependent statistical tilt symmetry with any
regulator of the form given in
Eqs.~(\ref{eq_masslike},\ref{eq_replicatedregulator}), which implies
that the $1$-replica part (first cumulant) of the effective average
action is unrenormalized along the flow. Similarly, the additional
inversion symmetries of the random anisotropy ($\vect \chi_a\cdot
\vect \chi_b \rightarrow -\vect \chi_a\cdot \vect \chi_b$) and the
random temperature ($\chi_a, \left\lbrace \chi_b\right\rbrace _{b\neq
  a}\rightarrow - \chi_a, \left\lbrace \chi_b\right\rbrace _{b\neq
  a}$) models are readily accounted for with the choice
$\widetilde{R}_k\equiv 0$. Truncation schemes naturally follow.

Taking into account the underlying supersymmetry that characterizes
the random field model for a gaussian distribution of the random
field\cite{parisi79} is much more involved. First, because one knows
that the supersymmetry, which goes with the dimensional reduction
property, must be broken in low enough dimension (at least, in $d=3$),
so that, even if the RG flow is started with an initial condition
obeying supersymmetry, a mechanism should be provided to describe a
spontaneous breaking of the supersymmetry. Secondly, the supersymmetry
shows up in a superfield formalism built with auxilliary fermionic and
bosonic fields, but it is far from transparent in the present
framework based on the fundamental fields. (This is true already at
the level of the initial condition of the RG flow.) We shall therefore
defer the proper resolution of this problem to a forthcoming
publication.\cite{tarjus07_susy} Note that an underlying supersymmetry
is also present in the random manifold model, where it also leads to
the $d\rightarrow d-2$ dimensional reduction. However, the pure model
with no disorder is merely a free field theory, and this is easily
accounted for.\cite{wiese05}

\subsection{Truncation schemes}

We have already stressed that solving the exact RG equation for the
effective average action requires approximations. The general
framework has proven quite versatile for devising efficient and
numerically tractable approximations which are able to describe both
universal and nonuniversal properties in any spatial dimension and to
capture genuine nonperturbative phenomena (see Introduction). Such
approximations generally amount to truncating the functional form of
the effective average action, which results in a self-consistent flow
that preserves the fundamental structure of the theory (as the
symmetries, see above).

If one is interested in the long-distance physics of a system and in
observables at small momenta, a systematic truncation scheme is
provided by the so-called ``\textit{derivative expansion}''.\cite{morris98b, berges02} It
consists in expanding the effective average action in increasing
number of derivatives of the field(s) and retaining only a limited
number of terms. The lowest order is the ``local potential
approximation'' (LPA)\cite{nicoll76} in which one only considers the flow of the
effective average potential, \textit{i.e.}, the effective average
action for a uniform field configuration. The field is not
renormalized and the associated anomalous dimension is equal to
zero. Field renormalization, which is important in the present problem
where one expects the anomalous dimension to be quite sizeable in low
dimensions (\textit{e.g.}, numerical estimates give $\eta \simeq0.5$
for the RFIM in $d=3$), requires to go beyond the LPA and to consider
the first order of the derivative expansion. Previous studies on a
variety of systems, including the pure $O(N)$ model, have shown that
the system's behavior is quantitatively very well described at this
level of approximation.\cite{morris95b,morris98d,berges02,delamotte03} Higher-order terms improve
the accuracy,\cite{canet03a,canet03b} but they rapidly become
untractable except in simple models.

For the disordered systems considered here, one more step is
needed. We have seen in section II-C that an \textit{expansion in
  number of free replica sums} can be used to generate the cumulants
of the renormalized disorder. Keeping only a limited number of terms
in the expansion therefore leads to a systematic truncation scheme. To describe both the thermodynamics and the
renormalized probability distribution of the disorder, one must
consider at least the first two cumulants, or equivalently, the second
order in the expansion in free replica sums.

Finally, on top of the two previous approximations, it may be useful,
and numerically more tractable, to expand the functions appearing in
the truncated effective average action in powers of the field
considered around a given (uniform) configuration. This configuration
can be taken either as zero everywhere or as a nontrivial
configuration that minimizes the effective average potential (here,
more precisely, its $1$-replica component that gives access to the
thermodynamics). Again, the accuracy and convergence properties of
such field expansions have been widely tested for many different
models. In the present case, and for reasons that will become clear
later on, field expansions should be used with great caution.

\subsection{Minimal truncation}

Given the general scheme presented above, the choice of a minimal
nonperturbative trucation is guided by a combination of factors:
experience gained from studies on other models, constraints associated
with the symmetries of the full theory, intuition or previous
knowledge concerning the physics of the problem at hand, requirement
of being able to recover as much as possible exact and perturbative
results in the appropriate limits, and of course, a practical
limitation coming with the numerical capability to actually solve the
set of RG flow equations.

As we have already alluded to, a description of the long-distance
physics of random field models and related disordered systems at least
requires to keep the first two cumulants of the disorder,
\textit{i.e.}, the first two terms, $\Gamma_{k,1}$ and $\Gamma_{k,2}$,
of the expansion of the effective average action in free replica
sums. Because of the anticipated nonnegligible value of the anomalous
dimension of the field $\eta$, one must also include in the
description at least the first order of the derivative expansion of
the first cumulant $\Gamma_{k,1}$. The resulting truncated functional
form of the effective average action then reads
\begin{equation}
\label{eq_truncation_Gamma}
\begin{split}
&\Gamma_k\left[\{\vect \phi_a\}\right ]=  \int_{\vect x}  \bigg\{\sum_{a=1}^n\bigg[U_{k}(\rho_a(\vect x)) +  \frac 1{2} Z_k(\rho_a(\vect x)) \vert \partial \vect \phi_a(\vect x)\vert ^2 \\&+ \frac 1{4} Y_k(\rho_a(\vect x)) (\partial \rho_a(\vect x))^2  \bigg] -\frac 1{2} \sum_{a,b=1}^nV_{k}(\vect \phi_a(\vect x), \vect \phi_b(\vect x))  \bigg\},
\end{split}
\end{equation}
where, as before, $\rho_a(\vect x)=\frac{1}{2}\vert \vect \phi_a(\vect
x)\vert^2$. In the above expressions, $U_k(\vect \phi_1) \equiv
U_k(\rho_1)$ is the effective average potential, which is equal to the
$1$-replica component $\Gamma_{k,1}$ evaluated for a uniform field and
will hereafter be simply denoted the $1$-replica potential; $V_k(\vect
\phi_1,\vect \phi_2)\equiv V_k(\rho_1, \rho_2,\vect \phi_1 \cdot \vect
\phi_2)$ is the $2$-replica potential and is equal to the $2$-replica
component $\Gamma_{k,2}$ evaluated for a uniform field
configuration. Physically, $U_k(\vect \phi_1)$ is a coarse-grained
Gibbs free energy and $V_k(\vect \phi_1,\vect \phi_2)$ is the second
cumulant of the renormalized disorder evaluated for uniform fields
(see Eqs.~(\ref{legendre_gamma_1}, \ref{eq_cumg2})). The two terms $
Z_k(\rho_1)$ and $Y_k(\rho_1)$ correspond to field renormalization
functions for the Goldstone and massive modes, respectively.

We note in passing that the fact that only the first two cumulants of
the disorder have been kept in the truncation does not imply that the
probability distribution of the renormalized disorder is actually
taken as gaussian. Indeed, as will be discussed in the companion
paper,\cite{tarjus07_2} the probability is not gaussian in
general. The truncation means that we have neglected the contribution
coming from the third cumulant in the RG flow of the second cumulant
and have therefore decoupled the hierarchy of flow equations for the
cumulants.

Being interested in the description of the models in the full $(N,d)$
diagram, we will have recourse to further approximations that make the
numerical resolution of the flow equations easier. More specifically,
we consider the lowest-order term of the field expansion of the field
renormalization functions around a nontrivial configuration,
$\rho_{m,k}=\frac{1}{2}\vert \vect \phi_{m,k}\vert^2$, which minimizes
the $1$-replica potential $U_k(\rho)$: $Y_k\equiv 0$ and $
Z_k(\rho)\equiv Z_{m,k}$, with $Z_{m,k}=Z_k(\rho_{m,k})$ and
$U'_k(\rho_{m,k})=0$. Physically, $\vect \phi_{m,k}$ is the
magnetization (order parameter) at the scale $k$. (If $\vect \phi_{m,k\rightarrow0}= 0$, the system is in an $O(N)$ symmetric phase whereas if $\vect \phi_{m,k\rightarrow0}\neq 0$, the system is in the phase with broken symmetry.)  $Z_{m,k}$ is chosen
as the field renormalization in the cutoff function
$\widehat{R}_{k}(q^2)$ (see Eq.(\ref{eq_scaledregulator})).

Finally, we simplify the resulting RG flow equations by setting the
off-diagonal cutoff function to zero, $\widetilde{R}_{k}\equiv 0$. As
will be shown, this choice leads in general to an explicit breaking of
dimensional reduction (despite the fact that the infrared regulators
vanish identically when $k\rightarrow0$). In the following
paper\cite{tarjus07_2} we shall discuss the way to nonetheless make
sense out of the results, the distinction between spurious and real
breaking of dimensional reduction being easily characterized. A
complete resolution of this issue will be provided when extending the
NP-FRG approach to the superfield formalism.\cite{tarjus07_susy}

With the above approximations which we shall refer to as the minimal
trucation, the self-consistent NP-FRG equations can be derived from
Eqs.~(\ref{eq_flow_Gamma1}--\ref{eq_flow_Gamma2_final}). The
flows of the $1$- and $2$-replica potentials read
\begin{equation}
\label{eq_flow_U}
\begin{split}
\partial_kU_{k}(\rho_1&)=
\dfrac{1}{2}  \int_{\vect q} \partial_k \widehat{R}_k(q^2) tr \bigg\{  \widehat{\vect P}_{k}^{[0]}(q^2;\rho_1) \\& - \widehat{\vect P}_{k}^{[0]}(q^2;\rho_1) V_{k}^{(11)}(\vect \phi_1 ,\vect \phi_1) \widehat{\vect P}_{k}^{[0]}(q^2;\rho_1)\bigg\} ,
\end{split}
\end{equation}

\begin{equation}
\label{eq_flow_V}
\begin{split}
\partial_k &V_{k}(\vect \phi_1,\vect \phi_2)= - \dfrac{1}{2}  \int_{\vect q} \partial_k \widehat{R}_k(q^2) tr \bigg\{  \widehat{\vect P}_{k}^{[0]}(q^2;\rho_1)\\& \bigg[ V_{k}^{(20)}(\vect \phi_1 ,\vect \phi_2)+ \widehat{\vect P}_{k}^{[0]}(q^2;\rho_2)V_{k}^{(11)}(\vect \phi_1 ,\vect \phi_2)^2 + \\& V_{k}^{(20)}(\vect \phi_1 ,\vect \phi_2) \widehat{\vect P}_{k}^{[0]}(q^2;\rho_1)V_{k}^{(11)}(\vect \phi_1 ,\vect \phi_1) + \\& V_{k}^{(11)}(\vect \phi_1 ,\vect \phi_1) \widehat{\vect P}_{k}^{[0]}(q^2;\rho_1)V_{k}^{(20)}(\vect \phi_1 ,\vect \phi_2) \bigg]  \widehat{\vect P}_{k}^{[0]}(q^2;\rho_2)\\& + perm (12)\bigg\} ,
\end{split}
\end{equation}
where the trace is over the $N$-vector components and, due to the $O(N)$ symmetry,  $V_{k}(\vect \phi_1 ,\vect \phi_2)\equiv V_k(\rho_1, \rho_2,z)$ with $z=\vect \phi_1 \cdot \vect \phi_2/\sqrt{4\rho_1 \rho_2}$; the (modified) propagator $\widehat{\vect P}_{k}^{[0]}(q^2;\rho)$ is given by 
\begin{equation}
\label{propag_O(N)}
\begin{split}
  \widehat{\vect
    P}_{k}^{[0]} (q^2;&\rho)^{\mu\nu}=\bigg[\dfrac{(1-\delta_{\mu1})}{Z_{m,k}q^2+\widehat{R}_k(q^2)+U'_k(\rho)}+
  \\&\dfrac{\delta_{\mu 1}}{Z_{m,k}q^2+\widehat{R}_k(q^2)+U'_k(\rho)
    +2\rho U''_k(\rho)}\bigg]\delta_{\mu\nu} ,
\end{split}
\end{equation}
 where  $\mu =1$ is chosen to be the direction of the field $\vect
  \phi$ and therefore corresponds to the massive mode while the
$(N-1)$ remaining components represent the Goldstone modes.

The flow of the field renormalization constant $Z_{m,k}$ is obtained
from the prescription
$Z_{k}(\rho)=\partial_{q^2}\Gamma_{k,1}^{(2)}(q^2;\rho)^{\mu\mu}\vert_{q^2=0}$
with $\mu$ chosen as a Goldstone mode ($\mu\neq 1$)\cite{berges02}
and from the condition $U'_k(\rho_{m,k})=0$. It can be explicitly
written as
\begin{equation}
\begin{split}
\label{eq_flow_Z}
  \partial_{k}Z_{m,k}= \partial_{q^2}\widetilde{\partial}_k
  \Bigg[4\includegraphics[width= .2
  \linewidth]{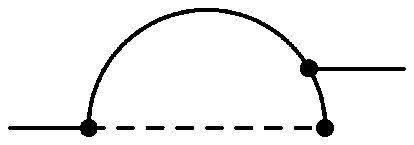}&-2\includegraphics[width=.2
  \linewidth]{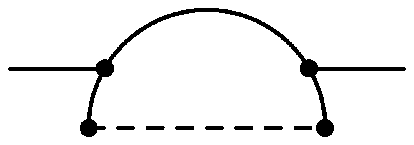}\\&- \includegraphics[origin=rb,width=.2
  \linewidth]{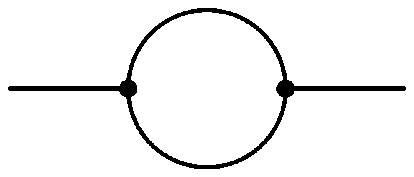} \Bigg]\Bigg |_{q=0} ,
\end{split}
\end{equation}
where a line denotes the Goldstone propagator and dots represent vertices obtained from derivatives of either the $1$-replica potential (single dots) or the $2$-replica potential (dots linked by a dashed line); for instance,
\begin{center}
 \includegraphics[width=.2\linewidth]{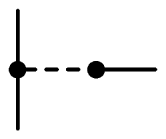}
\end{center}
represents the three-point vertex $\Gamma^{(21)}_k\equiv V_k^{(21)}$. We did not
include the graphs containing $4$-point vertices because in the
truncation considered here, they do not contribute to the
flow of $Z_{m,k}$. From the above flow equation, Eq.~(\ref{eq_flow_Z}), one extracts a running anomalous exponent,
\begin{equation}
\label{eq_def_running_eta}
\eta_k=-k \partial_{k}Z_{m,k}.
\end{equation}

The initial conditions for the RG flow equations are obtained from the
bare action, Eq.~(\ref{eq_replicated_generalaction}). The RG flow
equations form a closed set of coupled nonlinear integro-differential
equations for two functions, $U_{k}(\rho_1)$ and $V_k(\rho_1,
\rho_2,z)$, and a constant, $Z_{m,k}$. The numerical task of solving
these equations is still arduous and when needed for reducing the
difficulty of the computations, we will also consider truncated
expansions of the $1$- and $2$-replica potentials in some or all of
their field arguments (see below).

The present approach represents a nonperturbative but of course
approximate RG description. Already at the minimal truncation discussed
above, one includes all operators previously suggested to be important
for capturing the long-distance behavior of the present disordered
models, namely operators involving $1$- and $2$-replica terms. As will
be shown further below, it also reduces to the leading results of
perturbative RG analyses near the upper critical dimension,
$d_{uc}=6$, near the lower critical dimension for ferromagnetism when
$N>1$, $d=4$, and when the number of components $N$ becomes
infinite. One of its main advantages is that it provides a unified
framework to describe models in any spatial dimension $d$ and for any
number $N$ of field components. As such, it garantees a consistent
interpolation of all known results in the whole $(N,d)$ plane, in
addition to allowing the study of genuine nonperturbative
phenomena. If more accuracy is needed, the truncation scheme proposed
in III-B gives a systematic means to refine the description, by
including \textit{e.g.} the third cumulant or a more detailed account
of the momentum dependence of the $1-PI$ vertices.

In the following, we more specifically focus on the random field $O(N)$ model.

\section{Random field model}
\label{sec_RF}

\subsection{Scaling dimensions near a zero-temperature fixed point}

\begin{figure}[htbp]
  \centering
  \includegraphics[width=.8 \linewidth]{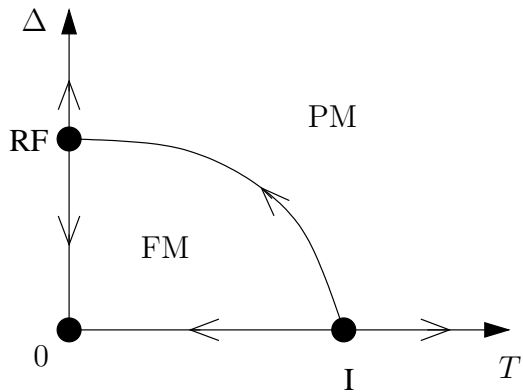}
  \caption{Schematic phase diagram of the RFIM in the disorder
    strength $\Delta$ - temperature $T$ plane above the lower critical
    dimension $d_{lc}=2$ (temperature can be
    introduced at the bare level through the Boltzmann weight). At low
    disorder and low temperature, the system is ferromagnetic, and it
    is paramagnetic otherwise. The arrows describe how the
    renormalized parameters evolve under the RG flow at long distance,
    and $I$ and $RF$ denote the critical fixed points of the pure and
    random-field Ising models, respectively. }
  \label{fig:phase_diag}
\end{figure}

For the RFIM, it has been proposed \cite{villain84,fisher86b}, and
convincingly supported by numerical and experimental results \cite{belanger98,nattermann98,middleton02}, that the fixed point controlling the critical
behavior associated with the transition between a high-temperature -
or large-disorder strength - disordered (paramagnetic) phase and a
low-temperature - or small-disorder strength - ordered (ferromagnetic)
phase is at zero temperature (see Figure \ref{fig:phase_diag}). 
The existence of such a
zero-temperature fixed point around which temperature is dangerously
irrelevant leads to a somewhat anomalous scaling at the critical
point.\cite{villain84,fisher86b} The two independent critical exponents characterizing the
scaling behavior of the pure Ising model should \textit{a priori} be
supplemented by an additional exponent $\theta$ describing the
vanishing of the (renormalized) temperature as the fixed point is
approached. This exponent $\theta$ leads to a modification of the
so-called hyperscaling relation, which becomes $2-\alpha = (d-\theta)
\nu$ where the critical exponents $\alpha$ and $\nu$ have their usual
meaning, and to a new scaling of the correlation functions. In
particular, the so-called ``connected'' and ``disconnected''
components of the pair correlation function (or $2$-point Green
function) behave at the critical point as:
\begin{equation}
\label{eq_Gconn}
G_{\text{c}}(q)=\overline{\langle\chi(- \vect q)\chi(\vect q)\rangle}-
\overline{\langle\chi(-\vect q)\rangle\langle\chi(\vect q)\rangle} \sim q^{-(2-\eta)}
\end{equation}
\begin{equation}
\label{eq_Gdisc}
G_{\text{d}}(q)=\overline{\langle\chi(-\vect q)\rangle\langle\chi(\vect q)\rangle} \sim q^{-(4-\bar\eta)}
\end{equation}
where $\eta$ is the usual anomalous dimension of the field and $\bar\eta$ is related to the temperature exponent $\theta$ through
\begin{equation}
  \label{eq_theta_eta_etabar}
\bar\eta =2-\theta+\eta.
\end{equation}

Above the upper critical dimension $d_{uc}=6$, the exponents take
their classical, mean-field values, $\eta=0, \alpha=0, \nu=1/2$, and
$\theta=2$, leading to $\bar \eta=0$. The dimensional reduction
property leads to a constant shift of dimension, $d\rightarrow d-2$,
\textit{i.e.}, to $\theta=2$ and $\bar\eta=\eta$, all exponents being
in addition given by those of the pure model in dimension
$d-2$. Whether the scaling behavior around the critical point is
described by $3$ independent exponents, or only $2$, has been a
long-time issue, with suggestions that an additional relation applies,
$\theta=2-\eta$ or equivalently $\bar\eta=2 \eta$.\cite{schwartz85b}
We shall address and answer this question in the following
paper.\cite{tarjus07_2}

To search for a zero-temperature fixed point, it is convenient to
introduce a renormalized temperature. Actually, one could add an
explicit temperature $T$ in the Landau-Ginzburg-Wilson description of
the model considered here: multiplying the argument of the exponential
in the partition function, Eq.~(\ref{eq_part_func}), by a factor
$T^{-1}$ to make the correspondence with the Boltzmann factor of
Statistical Physics leads to a bare replicated action in
Eqs.~(\ref{eq_replicated_bare}) and
(\ref{eq_replicated_generalaction}) in which the $1$-replica part,
including the kinetic term, is multiplied by a factor $T^{-1}$, the
$2$-replica part by $T^{-2}$, etc. Generally speaking, one can use
this temperature $T$ as a book keeping device to sort the orders
in the expansions in number of free replica sums. As a result for
instance, the modified propagator $\widehat {P}_{k}^{[0]}[\phi_1 ]$ is
proportional to $T$ whereas $\widetilde { P}_{k}^{[0]}[\phi_1, \phi_2
] $ is independent of $T$. One can use this book keeping trick to
devise ways to define a renormalized temperature at running scale $k$,
$T_k$, which reduces to the ``bare'' temperature $T$ at the
microscopic scale $k=\Lambda$. To this end, we first define the
renormalized disorder strength at scale $k$, $\Delta_{m,k}$, as
\begin{equation}
  \label{disstrength}
\Delta_{m,k} =\Delta_{k}\left(\phi_1=\phi_{m,k},  \phi_2=\phi_{m,k} \right),
\end{equation}
where as before $\phi_{m,k}$ is a field configuration that minimizes
the ($1$-replica) potential $U_k(\phi)$, and $\Delta_k(\phi_1,
\phi_2)$ is the second cumulant of the renormalized effective random
field defined as in Eq.~(\ref{eq_cumhren2}), namely,
\begin{equation}
  \label{Deltak}
\Delta_{k}\left(\phi_1,  \phi_2 \right)=V_k^{(11)}\left(\phi_1,  \phi_2 \right).
\end{equation}

In the present truncation, the second cumulant is only considered for homogeneous field configurations and $\Gamma_{k,2}^{(11)}$ reduces to $V_k^{(11)}$ with the same notations for partial derivatives as in Eqs.~(\ref{eq_notation_partderiv_W1}, \ref{eq_notation_partderiv_W2}) (\textit{e.g.}, $V_k^{(11)}\left(\phi_1,  \phi_2 \right)=\partial_{\phi_1} \partial_{\phi_2}V_k\left(\phi_1,  \phi_2 \right)$). At the microscopic scale $\Lambda$, $\Delta_{m,k}$ reduces to $\Delta_{\Lambda}/T^2$ where $\Delta_{\Lambda}$ is the bare variance of the random field and the factor $T^{-2}$ comes for reasons just explained above.

A running temperature can now be defined by
\begin{equation}
\label{eq_running_temperature}
T_k=\dfrac{Z_{m,k} \left( k/ \Lambda \right) ^2}{\left( \Delta_{m,k}/ \Delta_\Lambda \right) ^{2}}.
\end{equation}
One checks that since $Z_{m,\Lambda}=T^{-1}$ (see Eq.~(\ref{eq_replicated_generalaction}) and discussion above), $T_k$ indeed reduces to
$T$ when $k=\Lambda$. An associated running exponent
is obtained from
\begin{equation}
\label{eq_running_theta}
\theta_k=k \partial_k \ln T_k. 
\end{equation}
By using the definition of $\eta_k$, one may alternatively introduce a running exponent
$\bar{\eta}_k=2-\theta_k+\eta_k$, which converges to the critical exponent $\bar \eta$ defined in Eqs.~(\ref{eq_Gdisc}, \ref{eq_theta_eta_etabar}) if the relevant fixed point is reached, and compute it from the equation
\begin{equation}
\label{eq_running_etabar}
\bar{\eta}_k-2\eta_k= k\partial_k\Delta_{m,k}.
\end{equation}
On top of the usual scaling dimensions, $U_k,V_k \sim k^d$ and $\phi \sim (Z_{m,k}^{-1}k^{d-2})^{1/2}$, one can use the running temperature to define dimensionless quantities (denoted by lower-case letters) suitable for looking for a zero-temperature fixed point:
\begin{subequations}
  \label{al_dedim}
 \begin{align}
 &\phi=\left( \frac{k^{d-2}}{Z_{m,k} T_k}\right)^{1/2}\varphi\\&
 U_k\left(\phi_1 \right)=\frac{k^d}{T_k} u_k\left(\varphi_1\right)\\&
  V_k\left(\phi_1,  \phi_2 \right)=\frac{k^d}{T_k^2} v_k\left(\varphi_1,  \varphi_2 \right)\\&
  \Delta_k\left(\phi_1,  \phi_2 \right)=\frac{Z_{m,k} k^2}{T_k}\delta_k\left(\varphi_1,  \varphi_2 \right),
\end{align}
\end{subequations}
with $\delta_k\left(\varphi_1,  \varphi_2 \right)=v_k^{(11)}\left(\varphi_1,  \varphi_2 \right)$. Note that with the definitions of $\Delta_{m,k}$ and $T_k$, $\delta_{m,k}\equiv \delta_k\left(\varphi_{m,k},  \varphi_{m,k} \right)$  is constant along the RG flow and equal to its initial value $\Delta_{\Lambda}/\Lambda^{2}$ (in practice, and since we are not interested here in making a precise connection to the microscopic scale, we will set $\delta_{m,k}=1$).

\subsection{Scaled form of the exact RG equations for the RFIM}

With the use of the above defined dimensionless renormalized quantities, the flow equations can be expressed in a scaled form. Specifically, one can recast Eqs.(\ref{eq_flow_U}) and (\ref{eq_flow_V}) for $N=1$ in the form
\begin{equation}
\begin{split}
  \label{eq_u_ising}
\partial_t u_k& (\varphi)=-(d-2+\bar\eta_k-\eta_k)
u_k(\varphi)\\& + \frac{1}{2}(d-4+\bar\eta_k) \varphi
u_k'(\varphi)\\& +2v_d\left\{l_1^{(d)}(
u''_k(\varphi))\,\delta_k(\varphi,\varphi)+T_k l_0^{(d)}( u''_k(\varphi)) 
\right\}
\end{split}
\end{equation}
\begin{equation}
 \label{eq_v_ising}
\begin{split}
\partial_t v_k(\varphi_1&,\varphi_2)=-(d-4+2\bar\eta_k-2\eta_k)  v_k(\varphi_1,\varphi_2) + \\& \frac{1}{2}(d-4+\bar\eta_k) (\varphi_1\partial_{\varphi_1}+\varphi_2\partial_{\varphi_2}) v_k(\varphi_1,\varphi_2)\\& - 2v_d\Big\{
l_{1,1}^{(d)}(u_k''(\varphi_1),u_k''(\varphi_2))\delta_k(\varphi_1,\varphi_2)^2\\&+
l_{2}^{(d)}(u_k''(\varphi_1))\delta_k(\varphi_1,\varphi_1)v_k^{(20)}(\varphi_1,\varphi_2)\\& +
l_{2}^{(d)}(u_k''(\varphi_2))\delta_k(\varphi_2,\varphi_2)v_k^{(02)}(\varphi_1,\varphi_2)\\ & + T_k [l_{1}^{(d)}(u_k''(\varphi_1))v_k^{(20)}(\varphi_1,\varphi_2)\\& +
l_{1}^{(d)}(u_k''(\varphi_2))v_k^{(02)}(\varphi_1,\varphi_2) ] \Big\}
\end{split}
\end{equation}
where $\partial_t$ is a derivative with respect to $t=\ln(k/\Lambda)$, a prime
denotes a derivative with respect to the field (when only one argument is present), $v_d^{-1}=2^{d+1}\pi^{d/2} \Gamma(d/2)$, and we recall that $\delta_k(\varphi_1,\varphi_2)=v_k^{(11)}\left(\varphi_1,  \varphi_2 \right)$;  $l_{n}^{(d)}(w)$ and $l_{n_1,n_2}^{(d)}(w_1,w_2)$ are the ``dimensionless
threshold functions'' defined from the infrared cutoff function,
Eq.(\ref{eq_scaledregulator}), as :\cite{wetterich93,berges02}
\begin{equation}
  \label{eq_l_1}
  l_n^{(d)}(w)=-\frac 12(n+\delta_{n,0})\int_0^\infty
  dy\,y^{d/2}\frac{\eta_k\, r(y)+2 y r'(y)}{(p(y)+w)^{n+1}},
\end{equation}
\begin{equation}
\begin{split}
  \label{eq_l_2}
 & l_{n_1,n_2}^{(d)}(w_1,w_2)=-\frac 12 \int_0^\infty 
  dy\,y^{d/2}\left( \eta_k\,
  r(y)+2y r'(y)\right) \\& \frac{1}{(p(y)+w_1)^{n_1}(p(y)+w_2)^{n_2}} \left(\frac{n_1}{p(y)+w_1}+
  \frac{n_2}{p(y)+w_2}\right),
\end{split}
\end{equation}
with $p(y)=y(1+r(y))$ and $y=q^2/k^2$. The properties of these
threshold functions, whose detailed behavior depends on the choice of
the infrared cut-off function $r(y)$, have been extensively
discussed.\cite{wetterich93,berges02} They decay rapidly when $w\gg 1$, which,
since $u_k''(\varphi)=U_k''(\phi)/(Z_k k^2)$ is the square of a
renormalized mass, ensures that only modes with mass smaller than $k$
contribute to the flow in Eqs.~(\ref{eq_u_ising}) and
(\ref{eq_v_ising}). As an illustration, the use of the so-called
``optimized'' cut-off function
$r(y)=y^{-1}(1-y)\Theta(1-y)$,\cite{litim00} leads to explicit
expressions, namely,
\begin{equation}
  \label{eq_l_litim}
\begin{split}
l_{n_1,n_2}^{(d)}(w_1,w_2)=\frac {2}{d}&\left (1-\frac {\eta_k}{d+2}\right)\frac{1}{(1+w_1)^{n_1}(1+w_2)^{n_2}}\\& \times \left(\frac{n_1}{1+w_1}+
  \frac{n_2}{1+w_2}\right).
\end{split}
\end{equation}
The threshold functions essentially encode the nonperturbative effects
beyond the standard one-loop approximation. Note that, although not
shown in the notation, the threshold functions explicitly depend on
the scale $k$ via the running exponent $\eta_k$.

The above flow equations for $u_k(\varphi_1)$ and
$v_k(\varphi_1,\varphi_2)$ are supplemented by equations for $\eta_k$
and $\bar\eta_k$, \textit{i.e.}, for $Z_{m,k}$ and $T_k$ or
$\Delta_{m,k}$. (Note that the equation for $\bar\eta_k$ is actually
redundant as it is a consequence of the other equations; it is
nonetheless convenient to introduce and use it.) The flow equation for
$Z_{m,k}$ follows from Eq.(\ref{eq_flow_Z}) and one finds:
\begin{equation}
\begin{split}
  \label{eq_eta_ising}
\eta_k= \frac{4v_d}{d} & \Big\{4 m_{3,2}^{(d)} (u_{m,k}'',u_{m,k}'')u_{m,k}'''^2- \\&
2m_{2,2}^{(d)} (u_{m,k}'',u_{m,k}'') u_{m,k}''' \delta_{m,k}' \\& +
T_k m_{2,2}^{(d)} (u_{m,k}'',u_{m,k}'')u_{m,k}'''^2 \Big\}
\end{split}
\end{equation}
where we have used the short-hand notation
$\delta_k'(\varphi)\equiv \partial_\varphi \delta_k(\varphi,\varphi)=
\delta_k^{(10)}(\varphi,\varphi) + \delta_k^{(01)}(\varphi,\varphi)$
and the subscript ``$m,k$'' indicates that the functions are evaluated
for fields equal to $\varphi_{m,k}$; we have also introduced the
additional (dimensionless) threshold function
\begin{equation}
  \begin{split}
  \label{eq_threshold_m}
 & m_{n_1,n_2}^{(d)}(w_1,w_2)=-\frac12\int_0^\infty dy y^{
    d/2}\left( 1+r(y)+yr'(y)\right) \\&  \frac{1}{(p(y)+w_1)^{n_1}(p(y)+w_2)^{n_2}}
  \Bigg\{(1+r(y)+yr'(y))\\
  & (\eta_k r(y)+2yr'(y)) \left(\frac{n_1}{p(y)+w_1}+
    \frac{n_2}{p(y)+w_2}\right)-\\& 2\eta_k(r(y)+yr'(y))-
  4y(2r'(y)+yr''(y))\Bigg\},
      \end{split}
\end{equation}
whose properties are discussed in Ref.~[\onlinecite{wetterich93,berges02}]. For
instance, with the ``optimized'' regulator introduced above,\cite{litim00} one finds
that
\begin{equation}
  \label{eq_threshold_m-litim}
m_{n_1,n_2}^{(d)}(w_1,w_2)= \frac{1}{(1+w_1)^{n_1}(1+w_2)^{n_2}}.
\end{equation}

Finally the flow equation for $\Delta_{m,k}$ (or equivalently the flow
of the constraint $ \delta_{m,k}=1$ discussed below
Eq.~(\ref{al_dedim})) leads to the following equation:
\begin{equation}
  \label{eq_etabar_ising}
\begin{split}
2\eta_k- &\bar \eta_k=2v_d\bigg\{l_4^{(d)}(u''_{m,k}) u_{m,k}'''^2-
4 l_3^{(d)}(u''_{m,k}) u'''_{m,k} \delta'_{m,k}\\ &+l_2^{(d)}(u''_{m,k})(\delta''_{m,k}+\frac 32 \delta_{m,k}'^2-\frac{u'''_{m,k}}{u''_{m,k}}- \frac14 \Sigma_{m,k})\\& + l_1^{(d)}(u''_{m,k})
\frac{\delta_{m,k}'^2}{u''_{m,k}}-T_k \bigg[l_2^{(d)}(u''_{m,k}) u'''_{m,k}\delta'_{m,k}\\& - l_1^{(d)}(u''_{m,k})
(\frac 12 \delta''_{m,k}-\frac {u'''_{m,k}}{u''_{m,k}}\delta'_{m,k} +\frac 12 \widetilde\Sigma_{m,k})\bigg]\bigg\},
\end{split}
\end{equation}
where,  as before, $\delta_k'(\varphi)\equiv \partial_\varphi \delta_k(\varphi,\varphi)$ and similarly for $\delta_k''(\varphi)$, and we have  introduced 
\begin{equation}
  \label{eq_sigma}
\Sigma_k(\varphi_1)=\lim_{\varphi_2\to \varphi_1}(\partial_{\varphi_1}-\partial_{\varphi_2})^2
(\delta_k(\varphi_1,\varphi_2)-\delta_k(\varphi_1,\varphi_1))^2
\end{equation}
 and 
\begin{equation}
  \label{eq_tildesigma}
\widetilde\Sigma_k(\varphi_1)=
\lim_{\varphi_2\to \varphi_1}(\partial_{\varphi_1}-\partial_{\varphi_2})^2 \delta_k(\varphi_1,\varphi_2).
\end{equation}
All other notations are as before.

Before extending the results to the RF$O(N)$M, we point out important
features of the above equations. First, we have kept terms
proportional to $T_k$ but, provided one reaches a fixed point with an
exponent $\theta=\theta_{k\rightarrow 0}>0$ where temperature is thus
irrelevant, those terms are subdominant in the scaling region
$k\rightarrow 0$. In particular, the fixed point is attained by
following the flow with an initial temperature $T$ equal to zero.

Secondly, ``anomalous'' terms, $\Sigma_{m,k}$ and $T_k
\widetilde\Sigma_{m,k}$, appear in the expression of $2\eta_k- \bar
\eta_k$. As can be inferred from Eqs.~(\ref{eq_sigma}) and
(\ref{eq_tildesigma}), $\Sigma_{m,k}$ can only differ from zero, and
$\widetilde\Sigma_{m,k}$ become infinite, when a non-analyticity (a
``cusp'') in $(\varphi_1 - \varphi_2)$ appears in the (dimensionless)
renormalized disorder function $\delta_k(\varphi_1,\varphi_2)$ when
$\varphi_2\to \varphi_1$ (and both go to $\varphi_{m,k}$).  If
$\delta_k(\varphi_1,\varphi_2)$ is analytic, no signature of such
anomalous behavior is found. (We have implicitly assumed that no
stronger nonanalyticity appears, which means that a fixed point can be reached and that the theory is
renormalizable; this has to be checked in actual computations.) We
shall come back in more detail to these two important aspects of the
NP-FRG approach in the following paper.\cite{tarjus07_2} Finally, one may
notice that because of the $Z_2\equiv O(1)$ symmetry, the potential
$u_k$ is an even function of $\varphi$ and because of the additional
permutation symmetry,
$v_k(\varphi_1,\varphi_2)=v_k(\varphi_2,\varphi_1)=v_k(-\varphi_1,-\varphi_2)=v_k(-\varphi_2,-\varphi_1)$.

\subsection{Generalization to the RF$O(N)$M}

The preceding treatment can be extended to the RF$O(N)$M. The variable
$\rho=\frac{1}{2}\vert \vect \phi\vert^2$ is written in terms of a
dimensionless variable, $\widetilde\rho=k^{d-2}T_k^{-1} Z_{m,k}^{-1}\ \rho$ ,
where the tilde will be dropped in the following when no confusion is
possible between dimensionless and dimensionful quantities. The
variable $z=\vect \phi_1 \cdot \vect \phi_2/(2\sqrt{\rho_1\rho_2})$ is
already dimensionless.

For the $1$-replica second-order tensors (in $N$-vector components) evaluated for a uniform field configuration, \textit{e.g.}, for $\widehat{\vect P}_{k}^{[0]}(q^2; \vect \phi_1)$ or for $\Delta_k(\vect \phi_1,\vect \phi_1)\equiv V_k^{(11)}(\vect \phi_1,\vect \phi_1)$, the $O(N)$ symmetry reduces the number of terms to a ``longitudinal'' component (corresponding to the massive mode, see Eq.~(\ref{propag_O(N)})) and $N-1$ identical ``transverse'' components (corresponding to the Goldstone modes, see Eq.~(\ref{propag_O(N)})). We therefore introduce
\begin{equation}
\delta_k^{\mu\nu}(\rho,\rho,z=1)=\delta_{\mu\nu}\left[\delta_{\mu1}\delta_{k,L}(\rho)+(1-\delta_{\mu1})\delta_{k,T}(\rho) \right],
\end{equation}
with 
\begin{equation}
\label{eq_delta_kL}
\delta_{k,L}(\rho)=2\rho\partial_{\rho_1}\partial_{\rho_2}v(\rho_1,\rho_2,z=1)\vert_{\rho_1=\rho_2=\rho},
\end{equation}
\begin{equation}
\label{eq_delta_kT}
\delta_{k,T}(\rho) =\dfrac{1}{2\rho}\partial_{z}v(\rho,\rho,z)\vert_{z=1},
\end{equation}
and we define the longitudinal, $w_{k,L}(\rho)$, and transverse, $w_{k,T}(\rho)$, masses as
\begin{equation}
w_{k,L}(\rho)=u_k'(\rho)+2\rho u_k''(\rho),
\end{equation}
\begin{equation}
w_{k,T}(\rho)=u_k'(\rho),
\end{equation}
where a prime now denotes a derivative with respect to $\rho$.

The renormalized disorder strength at the running scale $k$ can be characterized, \textit{e.g.}, through the transverse component, $\Delta_{k,T}(\rho,\rho,z=1)$, evaluated for  $\rho=\rho_{m,k}=\frac{1}{2}\vert \vect \phi_{m,k}\vert^2$, and  $T_k$ is introduced accordingly. Expressing the $O(N)$ symmetry in the $2$-replica second-order tensors is a little more tedious, but nonetheless straighforward.

The resulting flow equations in scaled form read (where for ease of notation we drop the subscript $k$ in the right-hand sides, \textit{i.e.}, up to a sign, the beta functions, for all quantities but $T_k$ and also drop the argument of $v(\rho_1,\rho_2,z)$):
\begin{equation}
\label{eq_beta_uO(N)}
  \begin{split}
  \partial_t& u_k(\rho)= -( d - 2 + \bar \eta - \eta )u(\rho)+ \left( d-4 +\bar 
      \eta \right) \rho u'(\rho) \\&+ 2 v_d \bigg\{ (N-1)
      l_1^{(d)}(w_T(\rho))\delta_T(\rho)+l_1^{(d)}(w_L(\rho))\delta_L(\rho) \bigg\} \\&+ 2T_k v_d \bigg\{(N-1)
    l_0^{(d)}(w_T(\rho))+l_0^{(d)}(w_L(\rho)) \bigg\},
  \end{split}
\end{equation}
\begin{equation}
  \label{eq_beta_vO(N)}
  \begin{split}
&\partial_t  v_k(\rho_1,\rho_2,z)= -( d - 4 + 2\bar \eta  - 2 \eta) v+(d -4 + \bar \eta) \\& (
  \rho_1 \partial _{\rho_1} +\rho_2 \partial_{\rho_2})  v
%%%%%
- \dfrac{v_d}{4\rho_1\rho_2}\Big\{(N-1)\Big[4\rho_2 l_2^{d}(w_T(\rho_1))\\&\delta_T(\rho_1) (2\rho_1 \partial_{\rho_1} v- z \partial_z v ) + l_{1,1}^{(d)}(w_T(\rho_1),w_T(\rho_2))(\partial_z v)^2 \Big]\\& +(1-z^2)
%%%
\Big[4\rho_2 l_2^{(d)}(w_T(\rho_1))\delta_T(\rho_1) \partial_z^2 v +\\& 8\rho_2^2
l_{1,1}^{(d)}(w_T(\rho_1),w_L(\rho_2)) (\partial_{\rho_2}\partial_z v)^2 - l_{1,1}^{(d)}(w_T(\rho_1),w_T(\rho_2))\\&\big((\partial_z v)^2 
+2z\partial_z v\partial_z^2 v - (1-z^2) (\partial_z^2 v)^2\big) \Big]
%%%
\\& + 8 \rho_1 \rho_2 \Big[l_2^{(d)}(w_L(\rho_1)) \delta_L(\rho_1) (\partial_{\rho_1} v+ 2\rho_1\partial_{\rho_1}^2 v ) \\&+2 \rho_1 \rho_2 l_{1,1}^{(d)}(w_L(\rho_1),w_L(\rho_2)) (\partial_{\rho_1}\partial_{\rho_2}v)^2\Big]+perm(12)\Big\}
%%%
\\&- T_k \dfrac{v_d}{\rho_1\rho_2}
\Big\{ (N-1) \rho_2 l_1^{d}(w_T(\rho_1))( 2\rho_1\partial_{\rho_1} v -   z \partial_z v ) + \\&(1-z^2)\rho_2 l_1^{d}(w_T(\rho_1))\partial_z^2 v+
2\rho_1 \rho_2 l_1^{d}(w_L(\rho_1))\\&( \partial_{\rho_1} v+ 2\rho_1\partial_{\rho_1}^2 v) + perm(12) \Big\},
\end{split}
\end{equation}
\begin{equation}
  \label{eq_etaO(N)}
  \begin{split}
\eta_k=\dfrac{v_d}{d}\bigg\{&8 \bigg[m_{2,3}^{(d)}(w_{L}(\rho_m),0)+m_{3,2}^{(d)}(w_{L}(\rho_m),0)\bigg]\\& \delta_{T}(\rho_{m}) \dfrac{w_{L}(\rho_m)^2}{\rho_m}+ 8 m_{3,1}^{(d)}(w_{L}(\rho_m),0)\\& \times w_{L}(\rho_m)\bigg[\delta_{T}(\rho_{m})- \delta_{L}(\rho_{m})\bigg]\bigg\}
  \end{split}
\end{equation}
\begin{equation}
  \label{eq_etabO(N)}
\begin{split}
2\eta_k&- \bar \eta_k=\dfrac{2v_d}{\rho_{m}u''(\rho_{m})}\bigg\{(N-1)\rho_{m}l_1^{d}(0)\delta_{T}'(\rho_{m})^2+\\& l_2^{d}(0)u''(\rho_{m})+ l_2^{(d)}(w_L(\rho_m))\delta_{L}(\rho_{m})\bigg[(1+2\rho_{m}\delta_{T}'(\rho_{m})\\&+2\rho_{m}^2\delta_{T}''(\rho_{m}))u''(\rho_{m})- 2\rho_{m}^2 u'''(\rho_{m})\delta_{T}'(\rho_{m})\bigg]\\&-2 l_{1,1}^{(d)}(0,w_L(\rho_m))u''(\rho_{m})\bigg[1+\rho_{m}\delta_{T}'(\rho_{m})\bigg]^2\\&+\rho_{m}l_{1}^{(d)}(w_L(\rho_m))\delta_{T}'(\rho_{m})\delta_{L}'(\rho_{m})\bigg\}+ \cdots,
\end{split}
\end{equation}
where all symbols have the same meaning as in the previous equations
and, by construction, $w_{L}(\rho_m)=2\rho_m u''(\rho_m)$,
$w_{T}(\rho_m)=0$, and $\delta_{T}(\rho_{m})=1$. Note that in the last
two equations, we have omitted for simplicity the (subdominant) terms
involving $T_k$ in the beta functions and that in
Eq.~(\ref{eq_etabO(N)}), the dots denote ``anomalous'' terms which
generalize those found for the RFIM (see Eq.~(\ref{eq_etabar_ising}))
and vanish when the function $v_k(\rho_1,\rho_2,z)$ is analytic in all
its arguments; their expression is lengthy and will be discussed in
the companion paper.\cite{tarjus07_2}

When $N=1$ and $z=\pm1$, Eqs.~(\ref{eq_beta_uO(N)}) and
(\ref{eq_beta_vO(N)}) reduce to the previous equations for the RFIM,
Eqs.~(\ref{eq_u_ising}) and (\ref{eq_v_ising}), expressed with $\rho$
as variable instead of $\phi$: $v_k(\rho_1,\rho_2, z=+1)$ is equal to
$v_k(\varphi_1,\varphi_2)$ for $\varphi_1 \varphi_2 >0$ and
$v_k(\rho_1,\rho_2, z=-1)$ is equal to $v_k(\varphi_1,\varphi_2)$ for
$\varphi_1 \varphi_2 < 0$; $\delta_{k,L}(\rho)\equiv
\delta_{k}(\varphi)$ and $w_{k,L}(\rho)\equiv u''(\varphi)$.
\footnote{ Note that the anomalous dimension $\eta_k$ defined
  from the transverse (Goldstone) propagator in the RF$O(N)$M as in
  Eq.~(\ref{eq_etaO(N)}) and the exponent $\bar \eta_k$ defined from
  the transverse disorder strength as in Eq.~(\ref{eq_etabO(N)}) do
  not reduce to their RFIM counterparts given in
  Eqs.~(\ref{eq_eta_ising}) and (\ref{eq_etabar_ising}) when $N=1$,
  since the latter are obtained from longitudinal quantities. This
  results from the nonuniqueness of the prescription used for defining
  renormalized quantities in a truncated RG approach. This is already
  present in the pure $O(N)$ model\cite{berges02} and could be
  partly resolved by defining the anomalous dimension in the $O(N)$
  model from an expression involving some arithmetic mean of
  transverse and longitudinal components (and similarly for the
  disorder strength). In practice, one often chooses to keep for the
  Ising model the $O(N)$ expression with $N=1$. The ambiguity is more
  consistently resolved at the full first order of the derivative
  expansion\cite{canet03a}.} Finally, the comments made about
the important features of the flow equations for the RFIM carry over
to the equations for the RF$O(N)$M.

\subsection{Application to related disordered models}

Even though we have chosen to more specifically focus on the random
field model, it is worth sketching at this point the relevance of the
NP-FRG equations derived in this section to other disordered
systems. (As stressed already several times, we exclude spin glass
ordering from our considerations.)

The flow equations obtained for the RF$O(N)$M,
Eqs.~(\ref{eq_beta_uO(N)}-\ref{eq_etaO(N)}), directly apply to the
RA$O(N)$M for describing the long-distance physics associated with
ferromagnetic ordering. The putative fixed points are also expected to
be at zero temperature, so that similar scaling dimensions need be
introduced. The specificity of the random anisotropy model comes in
the initial conditions (see section II-A) and in the additional
symmetry of the $2$-replica potential, namely, $v_k(\rho_1,\rho_2,z)
=v_k(\rho_1,\rho_2,-z)$.

Similarly, the flow equations for the RFIM,
Eqs.~(\ref{eq_u_ising},\ref{eq_v_ising},\ref{eq_eta_ising}), can be
applied to the random elastic model. In this case, one can check that,
owing to the statistical tilt symmetry, $u_k'(\varphi)\equiv 0$ and
$\eta_k\equiv 0$ while $v_k(\varphi_1,\varphi_2)\equiv
v_k(\varphi_1-\varphi_2)$. After introducing the variable
$y=\varphi_1-\varphi_2$ and dropping the temperature,
Eq.~(\ref{eq_v_ising}), can be rewritten as
\begin{equation}
 \label{eq_v_elastic}
\begin{split}
-\partial_t v_k(y)=(d&-4+2\bar\eta_k) v_k(y) -\frac{1}{2}(d-4+\bar\eta_k) y v_k'(y)\\& + 2v_d
l_{2}^{(d)}(0)\left[ v_k''(y)-2v_k''(0)\right]  v_k''(y),
\end{split}
\end{equation}
where a prime denotes a derivative with respect to $y$.  The roughness
exponent is defined through $\zeta=-(d-4+\bar\eta)/2$, and one can
then see that the above equation reduces to the one-loop FRG equation
for a disordered elastic medium.\cite{fisher86,giamarchi98} Going beyond this level of description requires to
consider the next orders of the truncation scheme, in particular to
include the $3$-replica potential and apply the next order of the
derivative expansion for the $2$-replica effective average action.

Finally, Eqs.~(\ref{eq_u_ising},\ref{eq_v_ising},\ref{eq_eta_ising})
can be used in the case of the random temperature model with an
appropriate account of the symmetry: $u_k\equiv u_k(\rho)$, $v_k\equiv
v_k(\rho_1,\rho_2)$, with $\rho=\varphi^2/2$. However, the scaling
dimensions introduced to search for a zero-temperature fixed point are
not appropriate in the present case where one anticipates a fixed
point at a nonzero temperature (for a preliminary nonperturbative
treatment, see Ref.~[\onlinecite{tissier01b}]).

\section{Recovering the perturbative results}

\subsection{Analysis of the NP-FRG equations near $d=6$ and for $N \rightarrow \infty$}

For ease of notation, we only consider the RFIM, but a similar
analysis holds for the RF$O(N)$M.  It is easy to check that the flow
equations, Eqs.~(\ref{eq_u_ising},
\ref{eq_v_ising},\ref{eq_eta_ising},\ref{eq_etabar_ising}), admit for
fixed-point solution the Gaussian fixed point characterized by
$\eta_*^{(G)}=\bar\eta_*^{(G)}=0$,
$u_*^{(G)}(\varphi)=2v_dl_1^{(d)}(0)/(d-2)$, and
$\delta_*^{(G)}(\varphi_1, \varphi_2)=1$. The Gaussian fixed point is
once unstable for dimensions larger than $6$, but the coupling
constant associated with the $\varphi^4$-term in $u(\varphi)$ also
becomes relevant for dimensions less than $6$ so that the Gaussian
fixed point becomes unstable for $d<6$, as already well known.

The first order in $\epsilon=6-d$ can be derived by a direct expansion
of the fixed-point solution, with $u_*(\varphi)=u_*^{(G)}+\epsilon
u_1(\varphi)$, $\delta_*(\varphi_1,\varphi_2)=1+\epsilon
\delta_1(\varphi_1,\varphi_2)$. One easily finds that at this order
one still has $\eta_*=\bar\eta_*=0$. After inserting these results in
Eqs.~(\ref{eq_u_ising},\ref{eq_v_ising}), deriving the equation for
$v_k$ with respect to $\varphi_1$ and $\varphi_2$, and setting the
left-hand sides to zero, one obtains the following equations for
$u_1(\varphi)$ and $\delta_1(\varphi_1,\varphi_2)$:
\begin{equation}
  \label{eq_ising_gaussien_u}
\begin{split}
0=4u_1(\varphi)-\varphi u_1'(\varphi)-\frac{v_6}{2}\bigg[&l_1^{(6)}(0)-4l_2^{(6)}(0)
    u_1''(\varphi)\\& +4l_1^{(6)}(0)\delta_1(\varphi,\varphi)\bigg],
\end{split}
\end{equation}
\begin{equation}
\begin{split}
  \label{eq_ising_gaussien_v}
0=(\varphi_1\partial_{\varphi_1}+&\varphi_2\partial_{\varphi_2})\delta_1(\varphi_1,\varphi_2)
 \\& -2 v_6 l_2^{(6)}(0)
    (\partial_{\varphi_1}+\partial_{\varphi_2})^2 \delta_1(\varphi_1,\varphi_2) .
\end{split}
\end{equation}

By introducing the variables $x=(\varphi_1+\varphi_2)/2$ and  $y=(\varphi_1-\varphi_2)/2$, the latter equation can be rewritten as
\begin{equation}
  (x\partial_x+y\partial_y)\delta_1(x,y) = 2v_6 l_2^{(6)}(0) \partial_x^2 \delta_1(x,y).
\end{equation}
The symmetry of $v(\varphi_1,\varphi_2)$ with respect to the exchange of $\varphi_1$ and
$\varphi_2$ and to changes of sign of $\varphi_1$ and $\varphi_2$ (see IV-B) translates into the fact that $\delta_1$ is an even function of $x$ and $y$. Provided one requires that $\delta_1(x, y\to 0)$ is finite (which is needed for a well defined renormalizable theory), the only acceptable solution satisfying this property is a constant; due to the constraint $\delta_{m,k}\equiv \delta_{k}(\varphi_{m,k})=1$, it is equal to zero, \textit{i.e.}, $\delta_1(x,y)=0$. In addition to the now unstable Gaussian fixed point, Eq.~(\ref{eq_ising_gaussien_u}) has then for solution $u_1(\varphi)=(\lambda_{1*}/8)(\varphi^2-\varphi_{m*}^2)^2+ constant$ with $\varphi_{m*}^2=6v_6l_2^{(6)}(0)$. One also finds $\lambda_{1*}=(36 v_6 l_3^{(6)}(0))^{-1}$, so that, up to irrelevant constant factors, the solution corresponds to the fixed point of the pure Ising model (no random field) at first order in $\epsilon=4-d$. The fixed point is found once unstable and the associated exponents, \textit{e.g.}, $\nu=\frac{1}{2}+\frac{\epsilon}{12}$, satisfy the $d\to d-2$ dimensional reduction.

Equivalently, one can make a more direct connection to standard perturbation analysis by reframing the above results in a double expansion in $\epsilon$ and in the $\varphi^4$ coupling constant defined through $\lambda_k=u_k''''(\varphi_{m,k})$. Introducing as before $\rho_{m,k}=(1/2)\varphi_{m,k}^2$, one obtains from Eqs.~(\ref{eq_beta_uO(N)}-\ref{eq_etabO(N)}) that $\eta, \bar \eta = O(\lambda^2)$, $\delta=1+ O(\lambda^2)$ and
\begin{equation}
  \label{eq_ising_6-epslambda}
\partial_t \lambda_k= -\epsilon \lambda_k+ 36 v_d l_3^{(d)}(0)\lambda_k^2 + O(\lambda_k^3,T_k \lambda_k^2),
\end{equation}
\begin{equation}
  \label{eq_ising_6-epsrho}
\begin{split}
\partial_t \rho_{m,k}= -(2-\epsilon)  \rho_{m,k}+ 6 v_d \big[l_2^{(d)}&(0)-4l_3^{(d)}(0)\lambda_k  \rho_{m,k} \big] \\&+ O(\lambda_k^3,T_k \lambda_k^2),
 \end{split}
\end{equation}
where we have used the Taylor expansion of the threshold functions for
small arguments. (The fixed-point solution of
Eqs.~(\ref{eq_ising_6-epslambda},\ref{eq_ising_6-epsrho}) is of course
equal to that obtained above with $\lambda_*=\epsilon \lambda_{1*}$
and $\rho_{m*}=\varphi_{m*}^2/2$.) Again, up to irrelevant factors,
this gives back the one-loop perturbative result for the pure Ising
model obtained in a weak-coupling expansion in $d=4-\epsilon$.

The above result is derived through an expansion in a single coupling
constant, $\lambda_k$, associated to the $1$-replica part of the
effective action. It has been argued by Brezin and De
Dominicis\cite{brezin98,brezin01} that one should consider instead an expansion
involving all $\varphi^4$ coupling constants associated with multiple
replicas. In the present formalism, we can perform a more careful
analysis using the $\varphi^4$ coupling constants associated with the
$2$-replica part of the effective action, coupling constants that are
considered as potentially relevant in Refs.~[\onlinecite{brezin98,brezin01}]. We
find that this does not change the conclusion and, as previously
obtained in Ref.~[\onlinecite{mukaida04}], that the fixed point
corresponding to dimensional reduction is still once unstable at first
order in $\epsilon$. This is discussed in more detail in Appendix A.

The above analysis is extended to the $O(N)$ version in a
straightforward way. The property that the perturbative result at
first order in $\epsilon=6-d$ is recovered within our nonperturbative
approximation scheme is actually a consequence of the one-loop-like
structure of the exact flow equation for the effective average action,
Eq.~(\ref{eq_erg}). For the very same reason, the large $N$ limit can
also be easily recovered.

Rescaling the variables as $\rho \rightarrow N\rho$, $z\rightarrow z$
and the potentials as $u\rightarrow Nu$, $v\rightarrow Nv$, and
retaining only the dominant terms when $N\rightarrow \infty$, one
finds that $\eta=O(1/N)$, $\bar\eta=O(1/N)$ and that the
``longitudinal'' contributions drop out from the RG flow equations. As
a consequence, Eqs.~(\ref{eq_beta_uO(N)}) and (\ref{eq_beta_vO(N)})
can be recast as
\begin{equation}
\label{eq_beta_ulargeN}
 \begin{split}
\partial_t u_k(\rho)= -(d - 2)u_k(\rho)&+ (d-4) \rho u_k'(\rho)\\&+ 2 v_d l_1^{(d)}(u_k'(\rho))\delta_{k,T}(\rho),
  \end{split}
\end{equation}
\begin{equation}
\label{eq_beta_deltalargeN}
 \begin{split}
\partial_t& \delta_{k,T}(\rho_1,\rho_2,z)= (d-4)(\rho_1 \partial _{\rho_1} +\rho_2 \partial_{\rho_2})\delta_{k,T}(\rho_1,\rho_2,z) - \\&2 v_d\bigg\{\dfrac{l_{11}^{(d)}(u_k'(\rho_1),u_k'(\rho_2))}{\sqrt{\rho_1\rho_2}}\delta_{k,T}(\rho_1,\rho_2,z)\partial_z \delta_{k,T}(\rho_1,\rho_2,z) \\&+ \dfrac{l_2^{(d)}(u_k'(\rho_1))}{2\rho_1}\delta_{k,T}(\rho_1)(2\rho_1 \partial _{\rho_1}-z\partial_z) \delta_{k,T}(\rho_1,\rho_2,z) \\&+ \dfrac{l_2^{(d)}(u_k'(\rho_2))}{2\rho_2}\delta_{k,T}(\rho_2)(2\rho_2 \partial _{\rho_2}-z\partial_z) \delta_{k,T}(\rho_1,\rho_2,z)\bigg\},
  \end{split}
\end{equation}
where we have defined a generalized ``transverse'' disorder cumulant $\delta_{k,T}(\rho_1,\rho_2,z)$ via an extension of Eq.(\ref{eq_delta_kT}), namely,
\begin{equation}
\delta_{k,T}(\rho_1,\rho_2,z) =\dfrac{1}{2\sqrt{\rho_1\rho_2}}\partial_{z}v_k(\rho_1,\rho_2,z),
\end{equation}
which reduces to $\delta_{k,T}(\rho)$ when $\rho_1=\rho_2=\rho$ and $z=1$. Eq.~(\ref{eq_beta_deltalargeN}) is obtained by deriving the flow equation for $v_k(\rho_1,\rho_2,z)$.

If one starts the flow equations with an initial condition
$v_\Lambda(\rho_1,\rho_2,z)=2 \sqrt{\rho_1\rho_2}z$ (corresponding to
$\delta_{\Lambda,T}= 1$), the beta function is identically zero and
one therefore finds that the solution of
Eq.~(\ref{eq_beta_deltalargeN}) at all scales remains
$\delta_{k,T}(\rho_1,\rho_2,z)=1$. \footnote{ The same property does
  not hold for the RA$O(N)$M with \textit{e.g.} an initial condition
  $v_\Lambda(\rho_1,\rho_2,z)\propto \rho_1\rho_2z^2$, corresponding
  to $\delta_{\Lambda,T}\propto \sqrt{\rho_1\rho_2}z$. Actually, it is
  known from our previous study near $d=4$ that dimensional reduction
  is broken in this case: see Ref.~[\onlinecite{tissier06b}].} The
resulting equation for the $1$-replica potential is then very similar
to its counterpart for the pure $O(N)$ model with $N\rightarrow
\infty$ limit in dimension $d-2$ (the flow equation is then simply
given by the LPA \cite{berges02}).

To see more explicitly the connection, one can follow the flow of the $\varphi^4$ coupling constant $\lambda_k=u_k''(\rho_{m,k})$ as well as that of  $\rho_{m,k}$ which, we recall, satisfies $u_k'(\rho_{m,k})=0$ and is akin to a (dimensionless) order parameter at the running scale $k$. One finds
\begin{equation}
  \label{eq_lambdalargeN}
\begin{split}
\partial_t \lambda_k= -(6-& d) \lambda_k+ 4 v_d l_3^{(d)}(0)\lambda_k^2 \\&+ \big[(d-4)\rho_{m,k}-2v_d l_2^{(d)}(0)\big]u'''_k(\rho_{m,k}),
\end{split}
\end{equation}
\begin{equation}
  \label{eq_rholargeN}
%\begin{split}
\partial_t \rho_{m,k}= -(d-4)  \rho_{m,k}+ 2 v_d l_2^{(d)}(0),
%\end{split}
\end{equation}
which results in the nontrivial fixed point $\rho_{m*}= 2 v_d
l_2^{(d)}(0)/(d-4)$, $\lambda_*=(6-d)/(4 v_d l_3^{(d)}(0))$. This fixed point is once unstable (and it
remains so when considering the additional directions associated with
the $2$-replica potential, see above) and is characterized by critical
exponents satisfying the dimension reduction property, \textit{e.g.},
$\nu=1/(d-4)$ to be compared to $\nu=1/(d-2)$ for the pure model.

Note that the above perturbative expressions are recovered from the truncated NP-FRG equations even with an additional approximation using a field expansion around the minimum of the $1$-replica potential.

\subsection{Recovering the perturbative FRG near $D=4$}

A strong property of the minimal nonperturbative truncation described
above is that it also reduces, in the appropriate limit and for the
RF$O(N>1)$M, to the perturbative FRG equations at first order in
$\epsilon=d-4$ derived by Fisher.\cite{fisher85} The latter are
obtained from a low-disorder loop expansion of the nonlinear sigma
model associated with the RF$O(N)$M. It is therefore quite remarkable
that our formalism in which no hard constraint is enforced leads to
the proper result within the minimal approximation scheme.

For the RF$O(N)$M with $N>1$, $d=4$ is the lower critical dimension
for ferromagnetism. (We mean here long-range ferromagnetic order with a
nonzero order parameter, the case of quasi-long range order will be
discussed later on.) As a result, the critical point and the
associated fixed point occur near $d=4$ for a value of $\rho_{m}$ that
diverges as $1/\epsilon$ with $\epsilon=d-4$. As in the case of the
pure $O(N)$ model near $d=2$,\cite{delamotte03} one can therefore organize a systematic expansion in
powers of $1/\rho_{m}$.

At the minimum of the $1$-replica potential ($\rho=\rho_{m}$), the
transverse mass, associated with the Goldstone modes, is zero whereas
the longitudinal mass is very large and scales as $\rho_{m}$
(anticipating that $u''(\rho_{m})$ does not vanish). One can then use
the asymptotic properties of the threshold functions for large
arguments,
\begin{equation}
l_{n}^{(d)}(w\rightarrow \infty) \sim w^{-(n+1)}, \; l_{n_1,n_2}^{(d)}(w\rightarrow \infty,0)\sim  w^{-(n_1+1)},
\end{equation}
\begin{equation}
m_{n_1,n_2}^{(d)}(w\rightarrow \infty,0)\sim w^{-n_1},
\end{equation}
which encodes the decoupling of the massive mode.

In addition, we assume that as $\rho_{m}\rightarrow \infty$, $\delta_{L,T}(\rho_{m})$ stay finite (recall that actually, $\delta_{T}(\rho_{m})=1$) and that their derivatives, $\delta_{L,T}'(\rho_{m})$, etc, go to zero at least as fast as $1/\rho_{m}$; on the other hand, $\rho_{m}$ is a singular point for $u(\rho)$ (the location of its minimum), so that even when we expect that $u''(\rho_{m})$, $u'''(\rho_{m})$, etc, stay of $O(1)$. The consistency of these assumptions is easily checked \textit{a posteriori}. Inserting the above results and assumptions in Eqs.~(\ref{eq_etaO(N)}) gives
\begin{equation}
  \label{eq_eta_fisher}
\eta_k \simeq \dfrac{8v_d}{d\rho_{m,k}},
\end{equation}
which shows that $\eta$ is of order $1/\rho_{m}$.

Deriving once the flow equation for the $1$-replica potential $u_k(\rho)$ leads to
\begin{equation}
\label{eq_beta_u'_fisher}
 \begin{split}
\partial_t u'_k(\rho)= -(2&-\eta_k)u_k'(\rho)+ (\epsilon+\bar \eta_k) \rho u_k''(\rho)\\&- 2 v_d (N-1) l_2^{(d)}(u_k'(\rho))\delta_{k,T}(\rho),
  \end{split}
\end{equation}
from which one obtains the flow equation for the running order parameter $\rho_{m,k}$:
\begin{equation}
\label{eq_beta_rhom_fisher}
\partial_t \rho_{m,k}= -(\epsilon+\bar \eta_k) \rho_{m,k} + 2 v_d (N-1) l_2^{(d)}(0),
\end{equation}
where $\epsilon=d-4$. (Note that we have again omitted the subscript
$k$ in the right-hand sides and dropped the subdominant terms
involving the renormalized temperature $T_k$.) The last equation shows
that the fixed point value of $\rho_{m,k}$ satisfies, as anticipated,
$\rho_{m*}=O(1/\epsilon)$, which results in $\eta, \bar
\eta=O(\epsilon)$.

One can now apply a similar treatment to the flow equation for the
$2$-replica potential evaluated for $\rho_1=\rho_2=\rho_{m,k}$. For
convenience, we introduce the function
\begin{equation}
R_k(z) = \dfrac{v_k(\rho_{m,k},\rho_{m,k},z)}{(2\rho_{m,k})^2}
\end{equation}
which, due to Eq.~(\ref{eq_delta_kT}) and the constraint
$\delta_{k,T}(\rho_{m,k})=1$, satisfies $R'_k(z=1)=
1/(2\rho_{m,k})$. \footnote{ To make a more direct contact with the
  notations of Ref.~[\onlinecite{fisher85}], the function $R(z)$ used
  here differs by a factor $1/2$ from that used in our preceding
  Ref.~[\onlinecite{tarjus04}].} The flow equation for $R_k(z)$ can be
expressed as
\begin{equation}
\begin{split}
\partial_t R_k(z) = \dfrac{1}{(2\rho_{m,k})^2}&\partial_t v_k(\rho,\rho,z)\vert_{\rho=\rho_{m,k}}+\\&\partial_t \rho_{m,k} \;  \partial_\rho\left[\dfrac{v_k(\rho,\rho,z)}{(2\rho)^2} \right] \vert_{\rho=\rho_{m,k}},
\end{split}
\end{equation}
which with the help of Eq.~(\ref{eq_beta_rhom_fisher}) finally leads to
\begin{equation}
 \begin{split}
   \partial_t R_k(z)& \simeq (\epsilon+2 \eta_k) R(z) - 2 v_d
   l_2^{(d)}(0) \bigg\{(N-1) \bigg[\\&4R_k(z)R_k'(1)
   +R_k'(z)(R_k'(z)-2zR_k'(1))\bigg] +
   \\&(1-z^2)\bigg[-R_k'(z)^2+2(R_k'(1)-zR_k'(z))R_k''(z)\\&
   +(1-z^2)R_k''(z)^2\bigg]\bigg\}.
 \end{split}
\end{equation}
To dominant order in $\epsilon$, one can set $d=4$ in $v_d$ and 
$l_2^{(d)}(0)$ in all equations and in $v_d/d$ in
Eq.~(\ref{eq_eta_fisher}). Using the property of the threshold
function $l_2^{(d=4)}(0)=1+O(\eta)$ and discarding subdominant terms,
one finally arrives at
\begin{equation}
  \label{eq_eta_fisherfinal}
\eta_k = 4v_4R_k'(1), \; \bar \eta_k = -\epsilon + 4 (N-1) v_4 R_k'(1),
\end{equation}
\begin{equation}
  \label{eq_R(z)}
\begin{split}
  \partial_t R_k(z) =& \epsilon R_k(z) - 2 v_4 \bigg\{4(N-2) R_k'(1)
  R_k(z)+ \\& (N-1)\bigg[R_k'(z)-2zR_k'(1))\bigg]R_k'(z) \\&+
  (1-z^2)\bigg[-R_k'(z)^2  + 2(R_k'(1)-\\&zR_k'(z))R_k''(z)+
  (1-z^2)R_k''(z)^2\bigg]\bigg\},
 \end{split}
\end{equation}
where $v_4^{-1}=32\pi^2$ and $R_k(z)$ is of order $\epsilon$ near its
fixed point. The above equations coincide with the one-loop
perturbative FRG equations derived by Fisher.\cite{fisher85} Note that
this result is independent of the choice of the infrared cut-off
function $\widehat{R}_k(q^2)$: indeed, one easily checks that not only
$l_2^{(4)}(0)=1+O(\eta)$, but also $\lim_{w\rightarrow
  \infty}\big[m_{2,3}^{(4)}(w,0) w^2\big]=1+O(\eta)$, irrespective of
the regulator.

Finally, we note that setting $N=2$ and introducing the variable $\phi=cos^{-1}(z)$ in Eq.~(\ref{eq_R(z)}) leads to
\begin{equation}
 \label{eq_v_elastic_periodic}
%\begin{split}
-\partial_t R_k(\phi)=\epsilon R_k(\phi) - 2v_4 \left[ R_k''(\phi)-2R_k''(0)\right]  R_k''(\phi),
%\end{split}
\end{equation}
which, after use of Eq.~(\ref{eq_eta_fisherfinal}) for $\eta_k$ and
$\bar \eta_k$, coincides with the $1$-loop perturbative FRG equation
for a disordered periodic elastic system with a one-component
displacement field: compare for instance with
Eq.~(\ref{eq_v_elastic}), in which one should set $\zeta=0$ due to the
periodicity.\cite{giamarchi94} (Be careful, however, that
$\eta_k$ and $\bar \eta_k$ denote different sets of exponents in the
formalism leading to Eq.~(\ref{eq_v_elastic}) and in the present one.)
\footnote{ As will be discussed in more detail in the companion
  paper,\cite{tarjus07_2} the flow equations for the random elastic
  models are now considered for $d$ close to, but less than, $4$.}

\section{Concluding remarks}
\label{sec_conclusion}

In this work, described in the present paper and in the following
one,\cite{tarjus07_2} we have developed a theoretical approach which is
able to describe the long-distance physics, criticality, phase
ordering or ``quasi''-ordering, of systems in the presence of quenched
disorder, in particular random field models for which standard
perturbation theory is known to fail.

The approach is based on an exact renormalization group equation for
the effective average action (the generating functional of $1-PI$
vertices) and on a nonperturbative truncation scheme. This
nonperturbative RG formalism has recently been applied with success to
a variety of systems. The key point in the present problem is to
provide a proper account of the renormalized distribution of the
quenched disorder, and we have shown that this can be conveniently
done through a cumulant expansion and the use of a replica method in
which the permutational symmetry among replicas is explicitly broken.

We have stressed that any relevant treatment of random field models
and related disordered systems must include the second cumulant of the
renormalized disorder, \textit{i.e.}, at least a function of two
(replica) field arguments. Accordingly, we have proposed a
nonperturbative approximation scheme. Within this scheme, the minimal
truncation for the RF$O(N)$M already reproduces the leading results of
perturbative RG analyses near the upper critical dimension, $d_{uc}=6$
and when the number of components $N$ becomes infinite. More
importantly, it gives back the perturbative FRG equations near the
lower critical dimension for ferromagnetism when $N>1$, $d=4$.

One of the main advantages of the present approach, which will be
illustrated in the following paper, is that it provides a unified
framework to describe models in any spatial dimension $d$ and for any
number $N$ of field components. As such, it garantees a consistent
interpolation of all known results in the whole $(N,d)$ plane, in
addition to allowing the study of genuine nonperturbative phenomena.

We thank D. Mouhanna for helpful discussions.

\appendix
\section{Expansion in several coupling constants near $d=6$}

Near the upper critical dimension $d_{uc}=6$, the flow equations for
the RFIM derived within the minimal nonperturbative truncation,
Eqs.~(\ref{eq_u_ising}, \ref{eq_v_ising}, \ref{eq_eta_ising},
\ref{eq_etabar_ising}), can be expanded in several $\varphi^4$
coupling constants, in order to make the connection with recent
one-loop studies of the RFIM\cite{brezin98,brezin01,mukaida04}. On top of the
$1$-replica $\varphi^4$ coupling constant already used in section V-A,
$\lambda_k=u_k''''(\varphi_{m,k})$, we introduce two additional
coupling constants obtained from the $2$-replica potential,
\begin{equation}
  \label{2-replica_coupling_u_2}
u_{k,2}=-\dfrac{1}{2}\left(\partial_{\varphi_1}^2+\partial_{\varphi_2}^2 \right) \partial_{\varphi_1} \partial_{\varphi_2}v_k(\varphi_1,\varphi_2) \vert_{\varphi_1=\varphi_2=\varphi_{m,k}},
\end{equation}
\begin{equation}
  \label{2-replica_coupling_u_3}
u_{k,3}=-\partial_{\varphi_1}^2 \partial_{\varphi_2}^2 v_k(\varphi_1,\varphi_2) \vert_{\varphi_1=\varphi_2=\varphi_{m,k}},
\end{equation}
which amounts to consider a $2$-replica potential of the form
\begin{equation}
  \label{2-replica_potential}
\begin{split}
v_k(\varphi_1,\varphi_2)=\varphi_1 \varphi_2\bigg[& \delta_{m,k} - \dfrac{u_{k,2}}{6}\left(\varphi_1^2 +\varphi_2^2-6\varphi_{m,k}^2 \right)  - \\& \dfrac{u_{k,3}}{4}\left( \varphi_1 \varphi_2-4\varphi_{m,k}^2 \right)+ \cdots \bigg],
\end{split}
\end{equation}
where the dots denote higher-order terms in the field expansion around
the minimum of the $1$-replica potential and, as before,
$\delta_{m,k}=\partial_{\varphi_1} \partial_{\varphi_2}v_k(\varphi_1,\varphi_2)
\vert_{\varphi_{m,k}}\equiv 1$. The present description is thus very
similar to that used in Refs.~[\onlinecite{brezin98,brezin01}], except
that we do not include $3$- and $4$-replica terms. However, the issues
raised by Brezin and De Dominicis\cite{brezin98,brezin01} can already be
addressed by considering the $2$-replica term.

By expanding the flow equations for the $1$- and $2$-replica
potentials in powers of the coupling constants, which we generically
denote $u_{k,\alpha}$ with $u_{k,1}=\lambda_k$, one finds that $\eta =
O(u^2)$ and that up to a $O(u^3)$,
\begin{equation}
  \label{eq_brezin_g1}
\begin{split}
\partial_t  u_{k,1}= (d-6)& u_{k,1} + 2 v_d \bigg\{6 l_3^{(d)}(0) u_{k,1}^2+ 12 l_2^{(d)}(0) u_{k,1}\times \\&(u_{k,2}+u_{k,3})+  3 T_k l_2^{(d)}(0) u_{k,1}^2 \bigg\},
\end{split}
\end{equation}
\begin{equation}
  \label{eq_brezin_g2}
\begin{split}
\partial_t  u_{k,2}&= (d-4) u_{k,2} + 2 v_d \bigg\{6 l_3^{(d)}(0) u_{k,1}(u_{k,2}+u_{k,3})+ \\& 6 l_2^{(d)}(0) u_{k,2} (u_{k,2}+u_{k,3})+3 T_k  l_2^{(d)}(0) u_{k,1}u_{k,2} \bigg\},
\end{split}
\end{equation}
\begin{equation}
  \label{eq_brezin_g3}
\begin{split}
\partial_t  u_{k,3}&= (d-4) u_{k,3} + 2 v_d \bigg\{l_4^{(d)}(0) u_{k,1}^2 + 4 l_3^{(d)}(0) u_{k,1} \times \\&(u_{k,2} + u_{k,3})+  2 l_2^{(d)}(0) \left[ (u_{k,2}+u_{k,3})^2+3u_{k,3}^2\right] \\&+ 2 T_k  l_2^{(d)}(0) u_{k,1}u_{k,3} \bigg\},
\end{split}
\end{equation}
\begin{equation}
  \label{eq_brezin_lambda}
\begin{split}
  \partial_t &\left( \dfrac{\lambda_k \rho_{m,k}}{3}\right) = -2\left(
    \dfrac{\lambda_k \rho_{m,k}}{3}\right)+ 2 v_d \bigg
  \{l_2^{(d)}(0)u_{k,1}+ \\&2 l_1^{(d)}(0)
  (u_{k,2}+u_{k,3})+T_kl_1^{(d)}(0)u_{k,1}+\left( \dfrac{\lambda_k
      \rho_{m,k}}{3}\right)\times \\&\bigg[2
  l_3^{(d)}(0)u_{k,1}+8l_2^{(d)}(0)(u_{k,2}+u_{k,3})+T_kl_2^{(d)}(0)u_{k,1}\bigg]
  \bigg\},
\end{split}
\end{equation}
with, we recall, $ \rho_{m,k}=\varphi_{m,k}^2/2$. In addition, by
using $\partial_t T_k=(2+\eta_k-\bar \eta_k)T_k$ and the equation for
$2\eta_k-\bar \eta_k$, Eq.~(\ref{eq_etabar_ising}), one obtains to a
$O(u^2)$:
\begin{equation}
  \label{eq_brezin_temp}
\begin{split}
\partial_t T_k= 2 T_k& + 4 v_d  T_k \bigg \{2 l_2^{(d)}(0)(u_{k,2}+u_{k,3})+  6 l_1^{(d)}(0)\times \\& \frac{(u_{k,2}+u_{k,3})^2}{u_{k,1}} + T_k l_1^{(d)}(0)(2 u_{k,2}+3 u_{k,3})\bigg \}.
\end{split}
\end{equation}

It can now be checked that the above equations coincide with those
derived by Mukaida and Sakamoto \cite{mukaida04} with the introduction
of new running coupling constants: $g_0=T_k \delta_k(0)^{-1},
g_1=u_{k,1}\delta_k(0), g_2=u_{k,2}, g_3=u_{k,3}$, with
$\delta_k(0)\equiv \delta_k(\varphi_1=0,\varphi_2=0)= 1+
2(u_{k,2}+u_{k,3})\rho_{m,k}$ (and, of course, the $3$- and
$4$-replica contributions missing). As these authors, we therefore
obtain that the dimensional reduction fixed point is once unstable at
first order in $\epsilon=6-d$.

On the other hand, the analysis performed by Brezin and De Dominicis
\cite{brezin98,brezin01} requires to introduce different scaling dimensions,
corresponding to new coupling constants, $\hat{g}_0=g_0,
\hat{g}_1=g_1, \hat{g}_{2,3}=g_{2,3}g_0^{-1}$. The beta functions we
now obtain for $\hat{g}_0$, $\hat{g}_1$ and $\hat{g}_{2}$ coincide
with those of Ref.~[\onlinecite{brezin98,brezin01}], but that for $\hat{g}_3$
is ill defined as it contains a term that blows up as $k \rightarrow
0$. The scaling dimensions suggested by Brezin and De Dominicis are
thus not compatible with our approach.

\end{document}